\documentclass[prb,aps,showpacs,twocolumn,unsortedaddress,superscriptaddress]{revtex4}
\usepackage{graphicx}
\usepackage{float}
\usepackage{subfigure}
\usepackage{tabularx}
\usepackage{dcolumn}
\usepackage{amsbsy}
\renewcommand\vec[1]{\ensuremath\boldsymbol{#1}}
\newcommand\dquote[1]{\textquotedblleft #1\textquotedblright}

\begin{document}
\title{ Fermi surface topology and de Haas-van Alphen orbits in PuIn$_{\rm 3}$ and PuSn$_{\rm 3}$ compounds}

\author{C.-C. Joseph Wang}
\affiliation{Theoretical Division, Los Alamos National Laboratory, Los Alamos, New Mexico 87545, USA}
\author{M. D. Jones}
\affiliation{Department of Physics and Center for Computational Research, University at Buffalo-SUNY, Buffalo, New York 14260, USA}
\author{Jian-Xin Zhu}
\affiliation{Theoretical Division, Los Alamos National Laboratory, Los Alamos, New Mexico 87545, USA}
\date{\today}


\begin{abstract}
Since the recent discovery of plutonium-based superconductors such as PuCoIn$_{\rm 5}$, systematic studies
of the electronic properties for plutonium compounds are providing insight into the itinerancy-localization crossover of Pu 5$f$ electrons. We are particularly interested in understanding the Fermi surface properties of PuIn$_{3}$ compound, which serves as the building block for the PuCoIn$_{\rm 5}$ superconductor.
Motivated by the first observation of quantum oscillation and renewed interest in the de-Haas van-Alphen (dHvA) measurements on PuIn$_{\rm 3}$, we study the Fermi surface (FS) topology and
the band dispersion in both paramagnetic and antiferromagnetic state of  PuIn$_{\rm 3}$, based on density functional theory  with generalized gradient approximation.
We compare the results with its isostructural paramagnetic compound PuSn$_{\rm 3}$.
We present the detailed Fermi surfaces of compounds PuIn$_{\rm 3}$ and PuSn$_{\rm 3}$, and
conclude that the FS topology of an antiferromagnetic PuIn$_3$ agrees
better with dHvA measurements. In addition, we show the magnetization of the antiferromagnetic order can alter the field angle dependence and values of  the effective mass for the dHvA orbits. Our result suggest that  the accurate determination of the magnetic order orientation with respect to the crystal orientation is crucial
to advance the understanding of the electronic structure of the PuIn$_{\rm 3}$ compound.
\end{abstract}
\maketitle

\section{Introduction}
Actinide metals~\cite{Actinide1} are  strongly correlated electronic systems
due to the narrow bandwidth of $5f$ electrons.
In addition,  strong spin-orbit interaction within the $5f$ electron systems comparable to other energy scales  renders the understanding of the electronic properties of the actinide
metals more difficult.
Itinerant-to-localized crossover of 5$f$ electrons that occurs near plutonium in the actinide series is one of the most challenging issues in condensed matter physics, partly because
 the dual  character (partially localized/delocalized) of these 5$f$ electrons is closely related to the abrupt atomic volume variation between the $\alpha$-Pu   and $\delta$-Pu metals.
This change
in bonding leads to a 25\% larger volume in the $\delta$ phase as opposed to a low-symmetry, monoclinic crystal structure
$\alpha$ phase of Pu, along with a variety of unusual physical and mechanical properties.~\cite{Actinide2,Wang, Robert}

To understand the 5$f$ electron delocalization-localization crossover in elemental actinide solids,
it is very helpful to gain insight by studying its derivative compounds, which also show other emergent properties.
 One of the noticeable Pu-based compounds are 115 series.
It has been discovered recently that superconductivity occurs in PuCoGa$_{\rm 5}$~\cite{PuCoGa5} ($T_{c}=18.5$ K),  PuRhGa$_{\rm 5}$~\cite{PuRhGa5} ($T_{c}=8.7$ K),
and PuCoIn$_{\rm 5}$~\cite{PuCoIn5}
 ($T_{c}=2.5$ K) series.  The detailed pairing mechanism in Pu-115 compounds are currently under intense investigation.~\cite{Zhu,Oppeneer}
 We are particularly interested in the electronic properties of PuIn$_{\rm 3}$ and its isostructural partner PuSn$_{\rm 3}$.
 The reasons are twofold. First, the PuCoIn$_{\rm 5}$ structure consists of stacked CoIn$_{\rm 2}$ and  PuIn$_{\rm 3}$ layers,  therefore the understanding
 of the electronic properties in PuIn$_{\rm 3}$ is relevant to uncovering the mystery of superconductivity in  PuCoIn$_{\rm 5}$.
Second, the recent experimental capability to measure the Fermi surface topology of PuIn$_{\rm 3}$ by dHvA effects can help narrow down the minimal effective theory for these complicated systems.

In this paper,  we present systematic studies of the electronic structures on PuIn$_{\rm 3}$ and PuSn$_{\rm 3}$ within density functional theory (DFT).  We calculate the band dispersion, density of states (DOS), and the Fermi surface topology together with the identification of extremal dHvA orbits.
For PuIn$_3$, there is recent experimental evidence~\cite{PuIn3AFM} that the ground state of PuIn$_3$ is antiferromagnetic (AFM). However, the earlier dHvA measurement in this compound has been improperly interpreted in terms of  DFT calculations based on a paramagnetic (PM) state of PuIn$_3$.~\cite{HAGA} In the present work, we show the Fermi surface extremal orbits obtained for AFM PuIn$_{\rm 3}$  are in good agreement  with the dHvA measurements, which makes the theory and experiment comparison more
consistent. Also with the known experimental fact of PuSn$_{\rm 3}$ being paramagnetic in the ground state, we predict a quite different band structure and Fermi surface topology in PuSn$_{\rm 3}$ as opposed to PM PuIn$_{\rm 3}$. An experimental verification of this prediction provides strong supports for Landau's Fermi liquid theory and demonstrates the potential toward the detection of quantum oscillations in $\delta$-phase of Pu.

The organization of this paper is as follows.
In Sec. II,  we describe our theoretical method
and  briefly review the dHvA effects in itinerant electronic systems like metals.
In Sec. III A, we discuss the band structures
of the  PuIn$_{\rm 3}$ and PuSn$_{\rm 3}$ compounds, respectively.
In Sec. III B,  DOS for the two compounds in concert to
the band structures are discussed.
In Sec. III C, we show the Fermi surface topology and the corresponding dHvA  orbits,
which are relevant to available experimental studies.
Discussions and conclusions are summarized in section IV.
\begin{figure}
\includegraphics[scale=0.155]{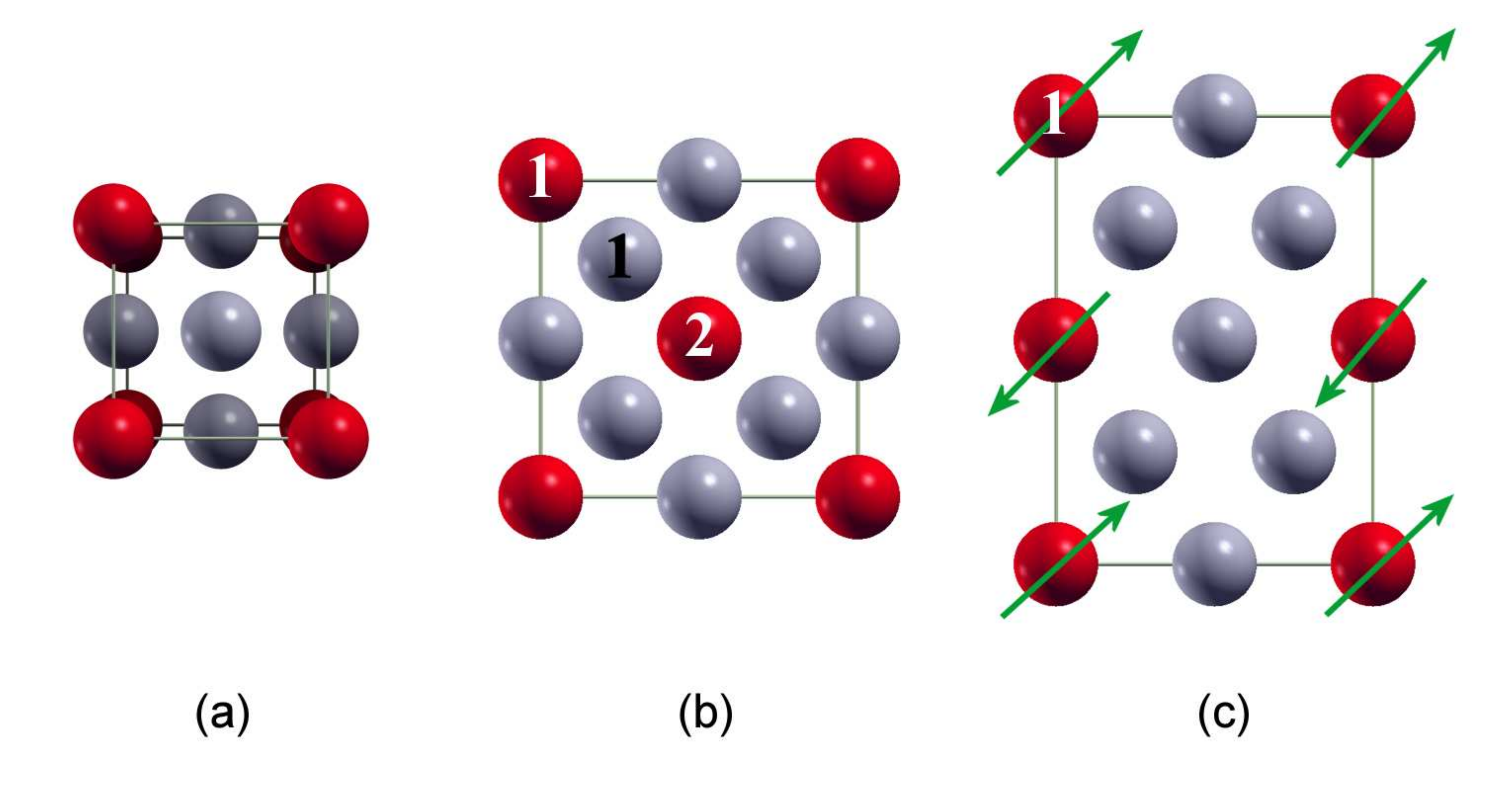}
\caption{(Color online) (a) PM PuM$_{\rm 3}$(M=In,Sn) conventional unit cell. Red sphere: Pu, Grey sphere: M;  (b) Top view of the anti-ferromagnetic PuIn$_{\rm 3}$ conventional unit cell; (c) Side view of  the AFM PuIn$_{\rm 3}$ conventional unit cell. The green arrows illustrates the commensurate AFM order.
The tilt arrow just indicates the spontaneous symmetry breaking orientation for viewing convenience. The actual orientation
for our studies can vary.  The AFM unit cell is enlarged twice with respect to PM unit cell  along the top-view axis  to accommodate the AFM oder.
The enlarged AFM unit cell coincides with two PM unit cells up to a 45-degree rotation with respect to the top-view axis.
}
\label{fig:FIG1}
\end{figure}

\section{Theoretical Method}
PuIn$_{{\rm 3}}$ and PuSn$_{{\rm 3}}$ compounds crystallize into
cubic AuCu$_{{\rm 3}}$-type structure at room temperature and have the actinide-actinide distance far above the Hill limit,~\cite{Hill} making the 5$f$-ligand
hybridization the dominant mechanism for Pu 5$f$-electron delocalization.
The experimental lattice constants at room temperature and atmosphere pressure ($4.61 \AA$  for PuIn$_3$ ~\cite{PuIn3} and  $4.63 \AA$  for PuSn$_3$ ~\cite{PuSn3}  respectively), are used in our calculation.
 PuSn$_{\rm 3}$ was reported as a paramagnet
experimentally.~\cite{PuSn3PM} However, PuIn$_{\rm 3}$ was originally assumed~\cite{HAGA} to be paramagnetic but more recent experiments indicate it is in an AFM state below 14 K.~\cite{PuIn3AFM} In this work,  we consider both PM and AFM cases for the PuIn$_{\rm 3}$ compound. The conventional unit cell of PM PuM$_{\rm 3}$
 (M=In, Sn)  is shown in Fig.~\ref{fig:FIG1}(a) with the ligand atoms M situated at face centers of the cell.  The magnetic unit cell for AFM PuIn$_{\rm 3}$ is enlarged because it is the commensurate AFM order obeying  the discrete translational invariance in the crystal.
Figure~\ref{fig:FIG1}(b) shows the top view of the magnetic unit cell accommodating the commensurate AFM order. The side view is shown in Fig.~\ref{fig:FIG1}(c).
Therefore, the lattices can be considered as a tetragonal Bravais lattice decorated with the basis of Pu and In atoms.

The first-principles calculations used here are based on the generalized gradient approximation with the PBE-96 functional.~\cite{PBE}
It has been well known that density functional theory under the local density approximation(LDA) underestimates
the correlation effects for late actinide metals including plutonium metals, leading to much smaller predicted atomic volume
with regards to experimental values.~\cite{Actinide1} With the generalized gradient approximation(GGA), the weak exchange correlation effects  are incorporated more reliably
than LDA. As a consequence, the GGA prediction generally moves the atomic volume closer to experimental values since the late actinide metals are on the strongly correlated side.
In this work, we are interested in plutonium-based  compounds (PuIn$_{\rm 3}$ and PuSn$_{\rm 3}$).
As suggested by the dominant role of majority ligand atoms  in other  Pu-115 compounds,~\cite{Zhu}
we confirm  the experimental equilibrium configuration is reliably determined by GGA studies with  the predicted atomic volumes well under three percent error regardless
of different magnetic orders.
Therefore, it is legitimate for us to use GGA approximation as a starting point to study the electronic properties of these plutonium compounds.

We use the scheme of the  full-potential linear augmented plane wave basis plus local basis~\cite{APW} (FP-LAPW+lo), as implemented in the WIEN2K code.~\cite{WIEN2K}
The LAPW sphere radii $R$ used for Pu, In, Sn atoms are 2.5 Bohr.
The interstitial plane wavenumber cut-off $k_{max }R=8.0$ is chosen for the  basis set.
The semi-core states (Pu $6s6p$, In $4p$) are included with the valence electrons using local orbitals. The core states are treated at the fully relativistic level with spin-orbital interaction for all atoms.
The spin-orbital interaction for semi-core and valence states are incorporated
by a second variational procedure~\cite{variation} using the scalar relativistic eigenstates as basis,  except that the so-called $6p_{1/2}$ relativistic local orbitals~\cite{phalf} are used to account for the finite character of the wave function at the Pu nucleus. Dense Brillouin zone sampling is used for the calculation of the Fermi surface.
We use a 31 $\times$ 31 $\times$ 31 k-mesh for the PM PuIn$_{\rm 3}$ and PuSn$_{\rm 3}$ calculations.
A 31$\times$ 31 $\times$ 22 k-mesh is used for the AFM PuIn$_{\rm 3}$.
Experimentally, a full determination of the magnetic structure for the AFM state of PuIn$_3$ is currently not available.
Therefore, we assume this structure follows that of its isostructural compound: CeIn$_3$.~\cite{CeIn1, CeIn2}  This choice is also supported by Fermi surface topology of PM PuIn$_{\rm 3}$ as discussed later.
The dHvA effect is observable in very clean metallic systems, typically in strong magnetic fields exceeding several tesla. On sweeping the magnetic field $\bf B$, one observes oscillations in the magnetization, which are periodic in an inverse magnetic field
due to the fact that the number of occupied Landau levels changes with the magnetic field.
The measurement of dHvA effect with varying magnetic field orientation is a powerful probe on Fermi surface topology in metallic and inter-metallic systems.
 The dHvA frequency $F$ in MKS units is related to the extremal Fermi surface cross-sectional area $A$ surrounded by extremal cyclotron
orbits perpendicular to the magnetic field orientation: \cite{dHva}
\begin{equation}
F=\frac{\hbar}{2\pi e}A
\end{equation}
where $e$ is the elementary charge.
In addition, the effective mass of electrons averaged around the cyclotron orbits can be determined by the damping strength of dHvA measurements as a function of temperature. Our determination of the dHvA orbits is based on the numerical algorithm
implemented by Rourke and Julian.~\cite{Julian}

\begin{figure}
\includegraphics[scale=0.6]{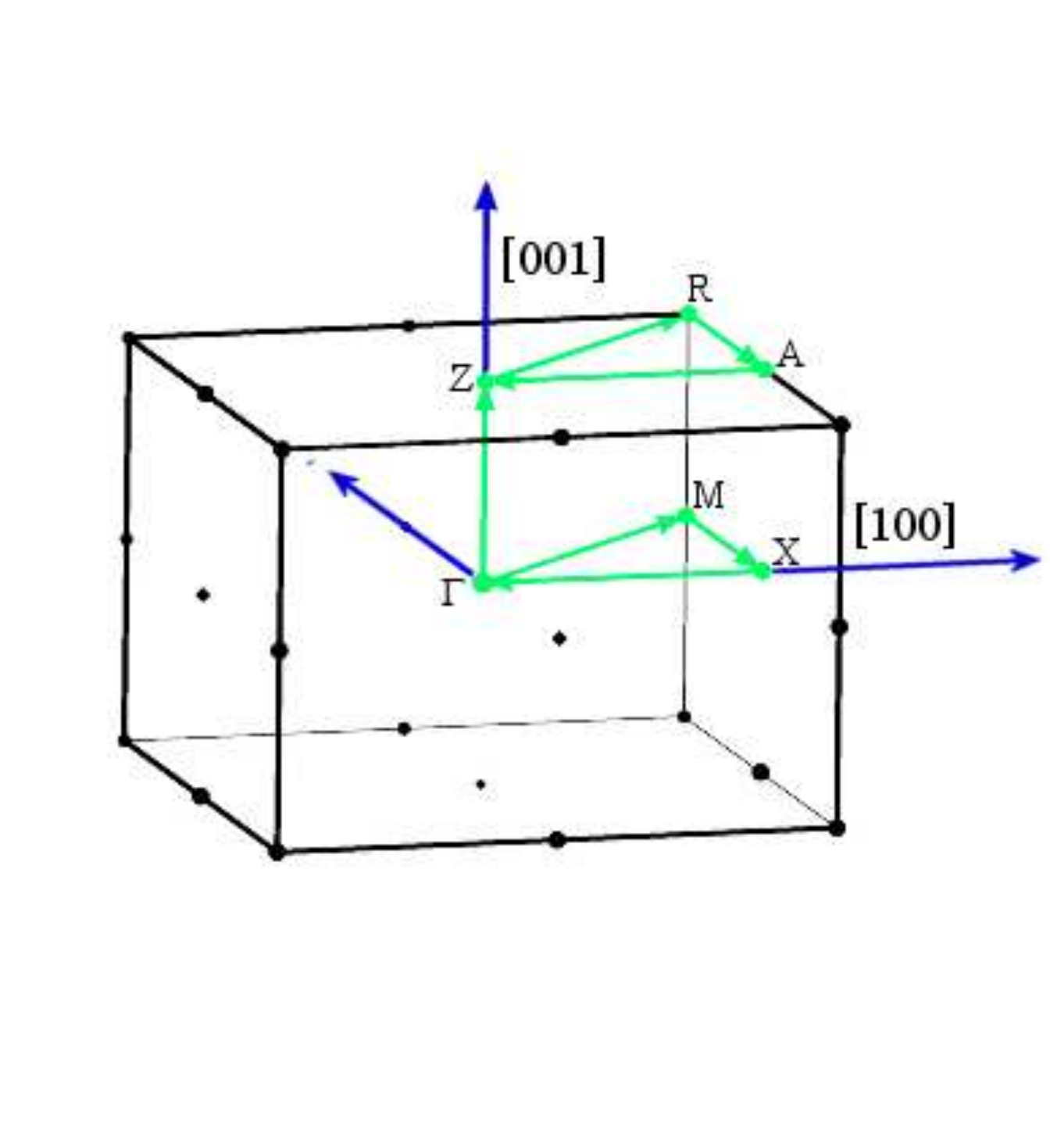}
\caption{(Color online) Tetragonal Brillouin zone (BZ). The symmetry points are labeled by capital letters. }
\label{fig:BZ}
\end{figure}

\begin{figure}
\includegraphics[scale=0.38]{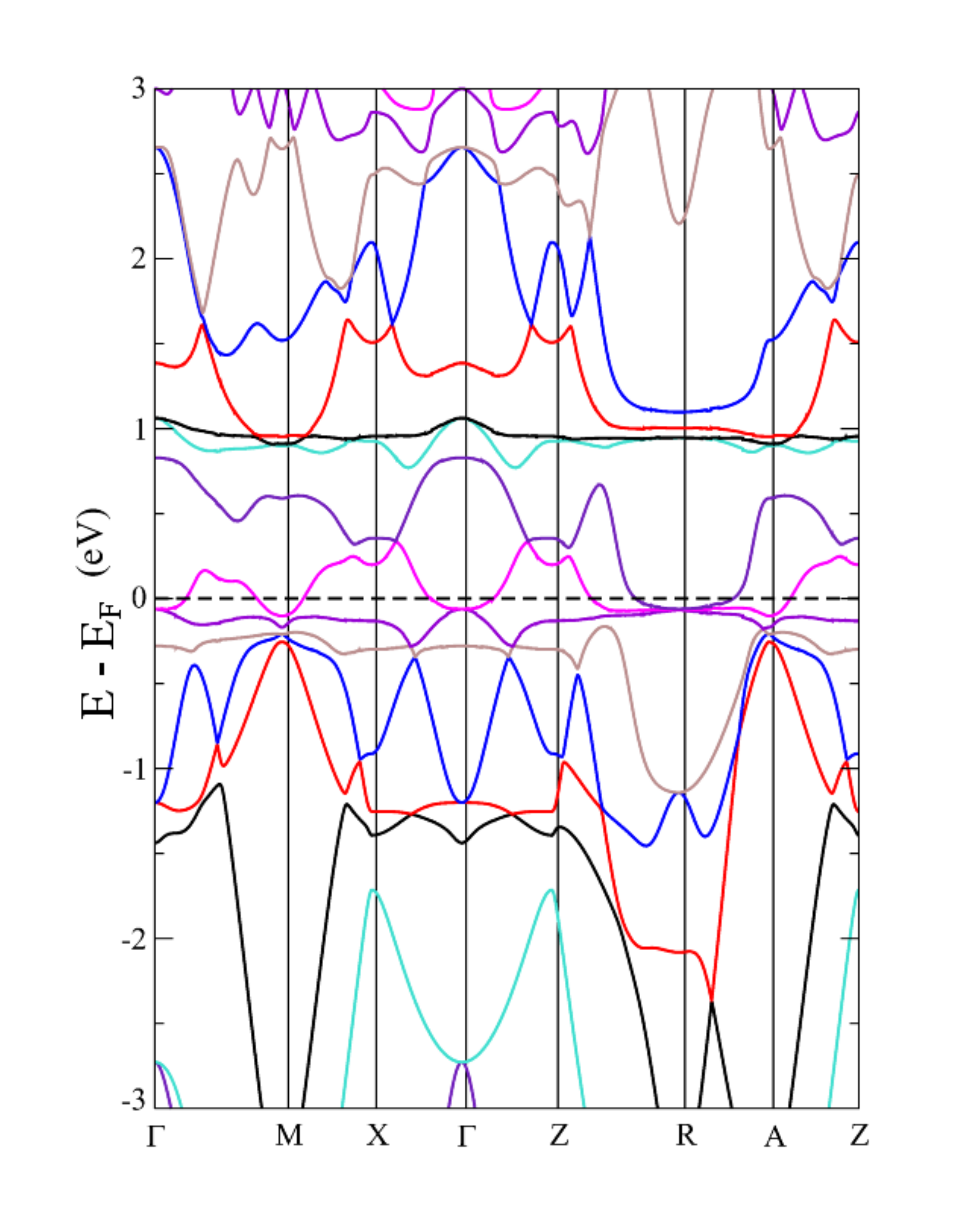}
\caption{(Color online) Band structure of the PM PuIn$_{\rm 3}$ along the high symmetric points in BZ. }
\label{fig:FIG_PuIn3_PMBand}
\end{figure}

\begin{figure}
\includegraphics[scale=0.38]{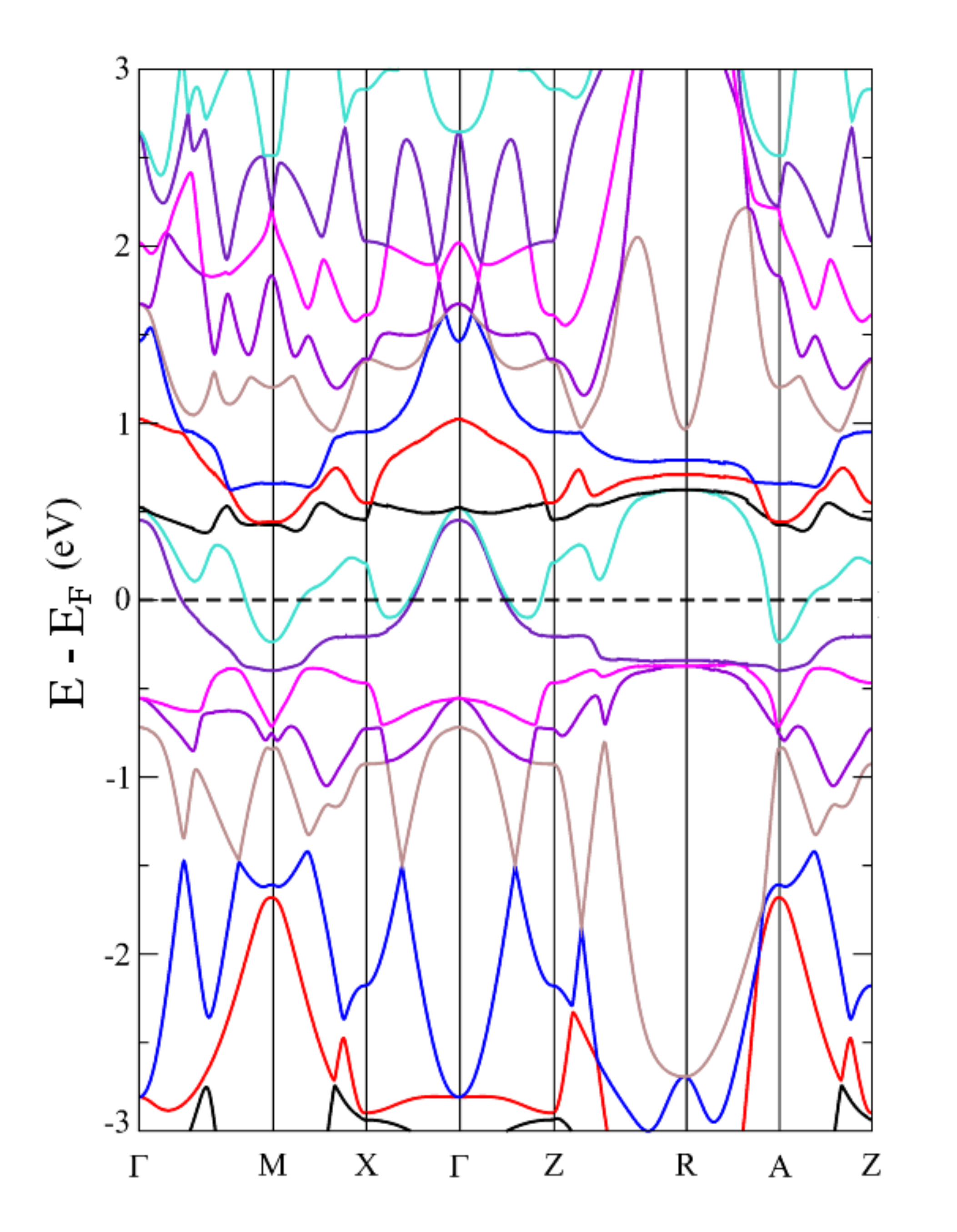}
\caption{(Color online) Band structure of the PM PuSn$_{\rm 3}$ along the high symmetric points in cubic BZ.
Two conduction bands are across the Fermi level $E_{F}$. }
\label{fig:FIG_PuSn3_Band}
\end{figure}

\begin{figure}
\includegraphics[scale=0.25]{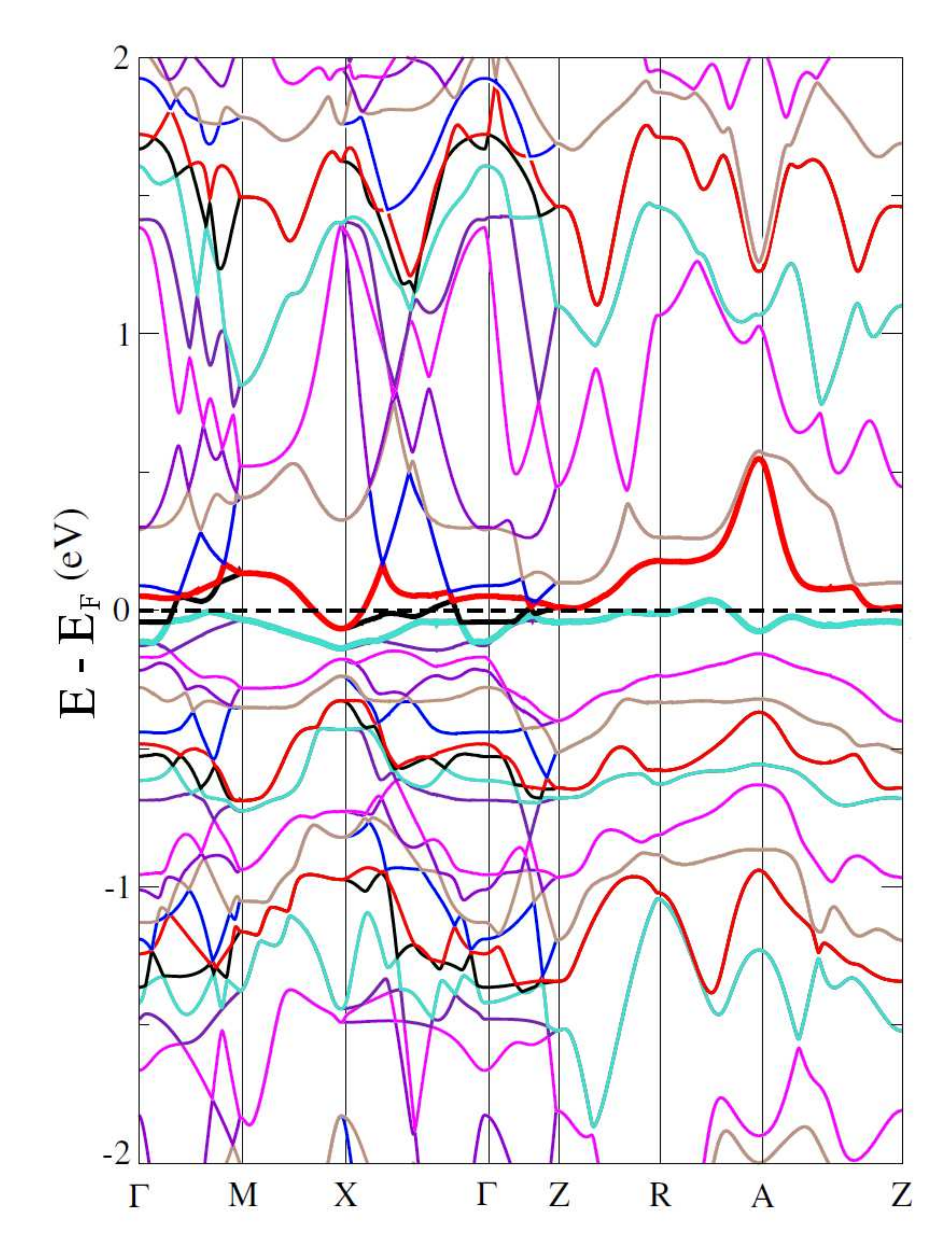}
\caption{(Color online) Band structure of the AFM PuIn$_{\rm 3}$ along the high symmetric points in tetragonal BZ  with the local spin polarization $\vec \sigma \parallel [001]$ .}
\label{fig:FIG_PuIn3_AFMBand}
\end{figure}

\section{Results}
\subsection{Band structures}
Since PM PuM$_{\rm 3}$ compounds can be generated from the cubic unit cell  in space, the corresponding BZ is cubic in reciprocal space.
For simplicity,  we label high-symmetry $\mathbf{k}$ points by the notation adopted by the tetragonal unit cell, as shown in Fig.~\ref{fig:BZ},  even though the BZ for the PM PuM$_{\rm 3}$ is cubic.

\begin{figure}[ht]
\centering
\includegraphics[scale=0.35]{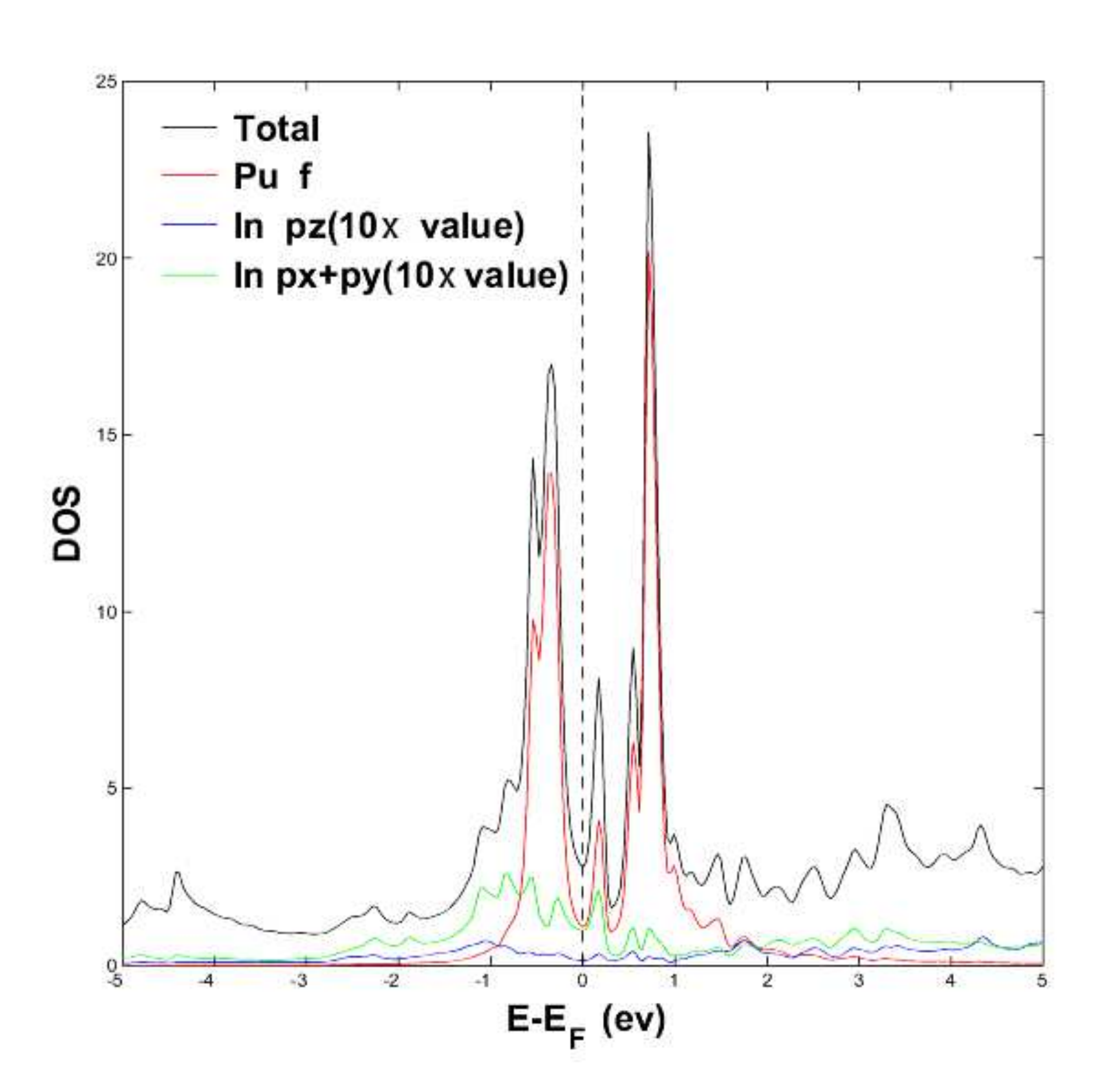}
\caption{(Color online)
DOS per atom and total DOS in PM PuIn$_{\rm 3}$ compound.
Ten times of the actual density of states for $p$ orbitals at In sites are plotted for better contrast.
The black solid line shows the total DOS in the BZ.
 }
\label{fig:FIG_PuIn3_PMDOS}
\end{figure}

The band structure for both  PM PuIn$_{\rm 3}$ and PuSn$_3$ demonstrates two bands across the Fermi energy. The energy  dispersion along the high
symmetry points is shown in Figs.~\ref{fig:FIG_PuIn3_PMBand} and~\ref{fig:FIG_PuSn3_Band}.  We observe the Fermi energy is crossed by two bands mainly of $5f$ electron character, as shown later in the density of states (DOS) studies.
The $f$-bands are far from flat  due to the hybridization mainly from the $5p$ valence electron of ligand atoms.
We calculate the band structure of PM PuIn$_{\rm 3}$ and PM PuSn$_{\rm 3}$ by the density functional theory using the GGA approximation.
We notice that the energy dispersion for PM PuSn$_{3}$ is very different, which indicates that the PM PuIn$_{\rm 3}$ and PM PuSn$_{\rm 3}$
have very different Fermi surface topology, as discussed in Sec. III C.

For the AFM PuIn$_{\rm 3}$ with the chosen magnetic unit cell,  the BZ in the reciprocal lattices becomes tetragonal.
Let us first discuss the circumstances in which magnetic moment orientation $\vec{\sigma}$ for the AFM order is
parallel to [001] direction defined by PM PuIn$_{\rm 3}$ crystal unit cell.
There are more atoms in the unit cell and therefore far more
bands are accommodated  in the corresponding tetragonal BZ. We observe four bands crossing the Fermi energy $E_{f}$ as shown in Fig.~\ref{fig:FIG_PuIn3_AFMBand}.
In this case, the bandwidth of $5f$ bands across Fermi surface is much narrower  than the bandwidth in the PM case.
The energy bands near the Fermi energy have dominant $5f$ electron features, as shown later in the DOS, and have large effective band mass
because of the flatness of the bands.

\begin{figure}[ht]
\centering
\includegraphics[scale=0.35]{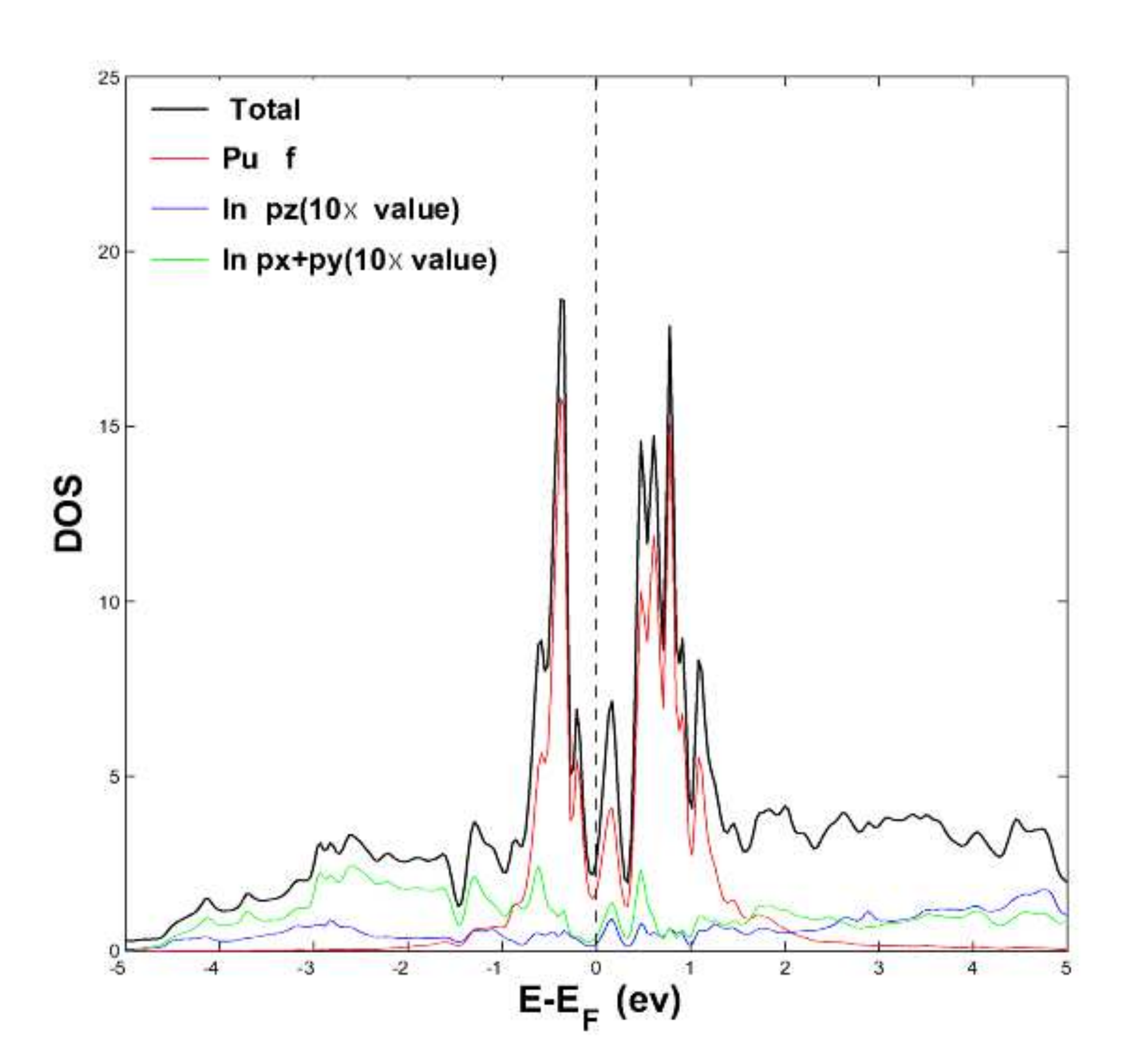}
\caption{(Color online) DOS per atom  and total DOS  in a PM PuSn$_{\rm 3}$ compound.
Ten times of the actual DOS for $p$ orbitals at Sn sites are plotted for better contrast.
The black solid line shows the total DOS in the BZ. }
\label{fig:FIG_PuSn3_PMDOS}
\end{figure}

\begin{figure}
\includegraphics[scale=0.35]{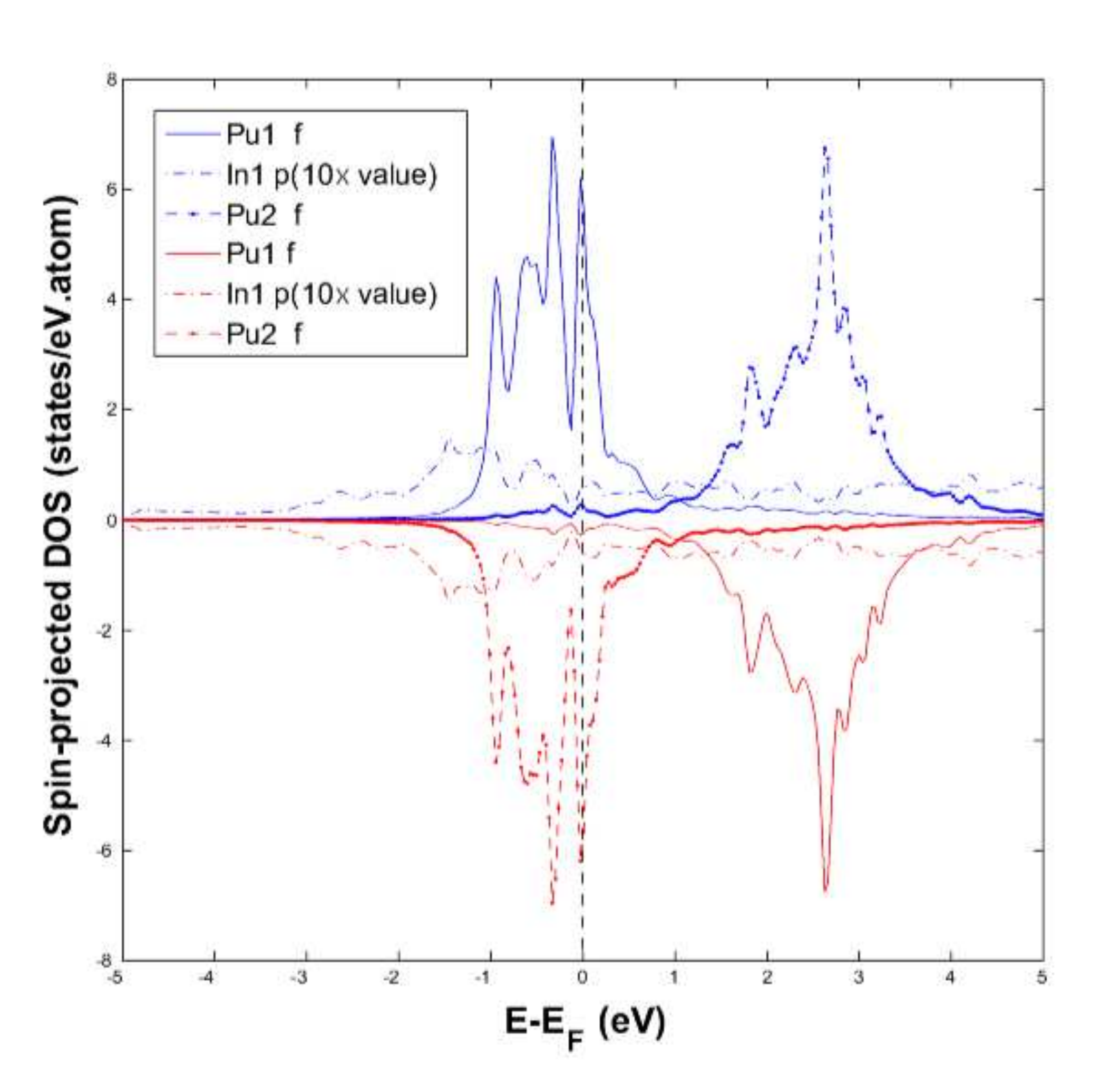}{l}
\caption{(Color online) Spin-projected DOS for AFM PuIn$_{\rm 3}$ compound.
Pu1 represents the Pu atom with a positive local magnetic moment.
Pu2 represents the Pu atom with a negative local magnetic moment.
It is noted that Pu1 atom and Pu2 atom are the nearest neighbors in the AFM unit cell.
The In1 atom indicates the indium atom located between the Pu1 and Pu2 atom in the unit cell (Figs.~1(b) and 1(c)).
The color \dquote{blue} indicates the  up-spin projected DOS along the chosen quantization $+z$ axis in Bloch sphere.
The color \dquote{red} indicates the down-spin projected DOS along the $-z$ axis in Bloch sphere.
Ten times of the actual spin-projected DOS of $p$ orbitals at the In1 site are plotted for better contrast. }
\label{fig:FIG_PuIn3_AFMDOS}
\end{figure}

\subsection{Density of states}

By the  smaller equilibrium volume  of PuIn$_{\rm 3}$ in comparison with PuSn$_{\rm 3}$,
one may argue that PM PuIn$_{\rm 3}$ should have a more itinerant $5f$ bands in comparison with
PuSn$_{\rm 3}$ due to the smaller distance between nearest-neighboring Pu atoms and the In atoms.
In addition, the atomic $5p$ levels from ligand atoms in PuIn$_{\rm 3}$ tend to be more delocalized  due to smaller attraction from nucleon charges.
However, those effects are not expected to be very remarkable in the DOS because of the closeness of the atomic properties for the In atom  and Sn atom.

To understand the hybridization between Pu 5$f$ and ligand valence states, we perform the DOS calculation for these compounds.
In Fig.~\ref{fig:FIG_PuIn3_PMDOS}, the DOS for the PM PuIn$_{\rm 3}$ is shown.
We observe a nonzero total DOS near the Fermi level.
The peak near the Fermi energy shows a hybridization of Pu 5$f$
with the DOS of $5p$ valence electrons from the In atom. This indicates that the metallic behavior is inherent from the hybridization of  $5f$ localized electrons at Pu with $5p$ itinerant electrons at In. The two strong peaks away from the Fermi surface are due to the strong spin-orbital coupling for the localized $f$ electrons, which is weakly hybridized
with In $5p$ orbitals. The lower peak is mainly due to $j=5/2$ sub-bands  and the higher peak  is mainly due to $j=7/2$ sub-bands, which has been studied previously by relativistic Stoner theory.~\cite{Olle Erickson}
The $5f$ electron occupancy at Pu atomic sphere is around 4.8. This suggests that the mid-peak  near Fermi energy
originates from the higher $j=5/2$ sub-bands in the presence of crystal-field splitting.
The three-peak features are similar also in PuSn$_{\rm 3}$ except
the $j=5/2$ peak are broadened by  the hybridization from Sn $p$ orbitals.
Fig.~\ref{fig:FIG_PuSn3_PMDOS} presents the DOS for PuSn$_{3}$. We notice the $j=5/2$ sub-bands are still sharply peaked
but with a slightly larger $5f$-DOS at Pu site as shown by the solid-red line as suggested by a slightly smaller hybridization.
As will be discussed later, PM PuIn$_{\rm 3}$ and PM PuSn$_{\rm 3}$ have very
different Fermi surface topology despite the similarity in DOS.

We have also calculated the DOS for AFM PuIn$_3$, as shown in Figs.~1(b) and 1(c), the lattice can be divided into two sub-lattices. The calculated spin projected partial DOS  at different atomic sites is shown in Fig.~\ref{fig:FIG_PuIn3_AFMDOS}.
We observe that the Pu1 atom has larger 5$f$ occupancy in the up-spin channel (see corresponding partial DOS as denoted by ``blue'' solid line). However, the Pu2 atom takes larger 5$f$ occupancy in the down-spin channel (see corresponding partial DOS as denoted by ``blue'' dotted line).
This indicates that Pu1 and Pu2 atoms have magnetic moments due to $5f$ electrons, which are equal in magnitude but antiparallel in orientation.
We do not observe net magnetic moment at In atoms according to the equal spin-projected DOS for the In1 atom indicated by the solid-dashed curves.
The AFM order has the effect of generating a gap, above the Fermi energy, between two major peaks belonging to up and down spin Pu 5$f$-DOS.
By our first-principle calculation, we found the orbital moment at Pu1 atom is around
$-2.1 \mu_{B}$ ($\mu_{B}:$ Bohr magneton) but changes sign (-$2.1\mu_{B}$) for Pu2 atom. The spin moment at Pu1 and Pu2  atoms are given by
 $4.7\mu_{B}$ and $-4.7\mu_{B}$ respectively.
We observe the presence of magnetic order causes the mixing of $j=5/2$ and $j=7/2$ and turn the local dip structure near the Fermi surface to
a peak structure.  In addition, two major peaks in DOS  appear away from Fermi surface with large energy separation between maxima.

\subsection{Fermi surface topology and dHvA orbits}

In this section, we show the Fermi surface topology and calculate the dHvA frequencies as a function of magnetic field
orientation for compounds PuSn$_{\rm 3}$ and PuIn$_{\rm 3}$.  By the band structure calculation,
Fermi surface topology is determined by the isosurface calculated by the eigenstates of Kohn-Sham equation.

\subsubsection*{PM PuIn$_{\rm 3}$ compound}

\begin{figure}
\includegraphics[scale=0.4]{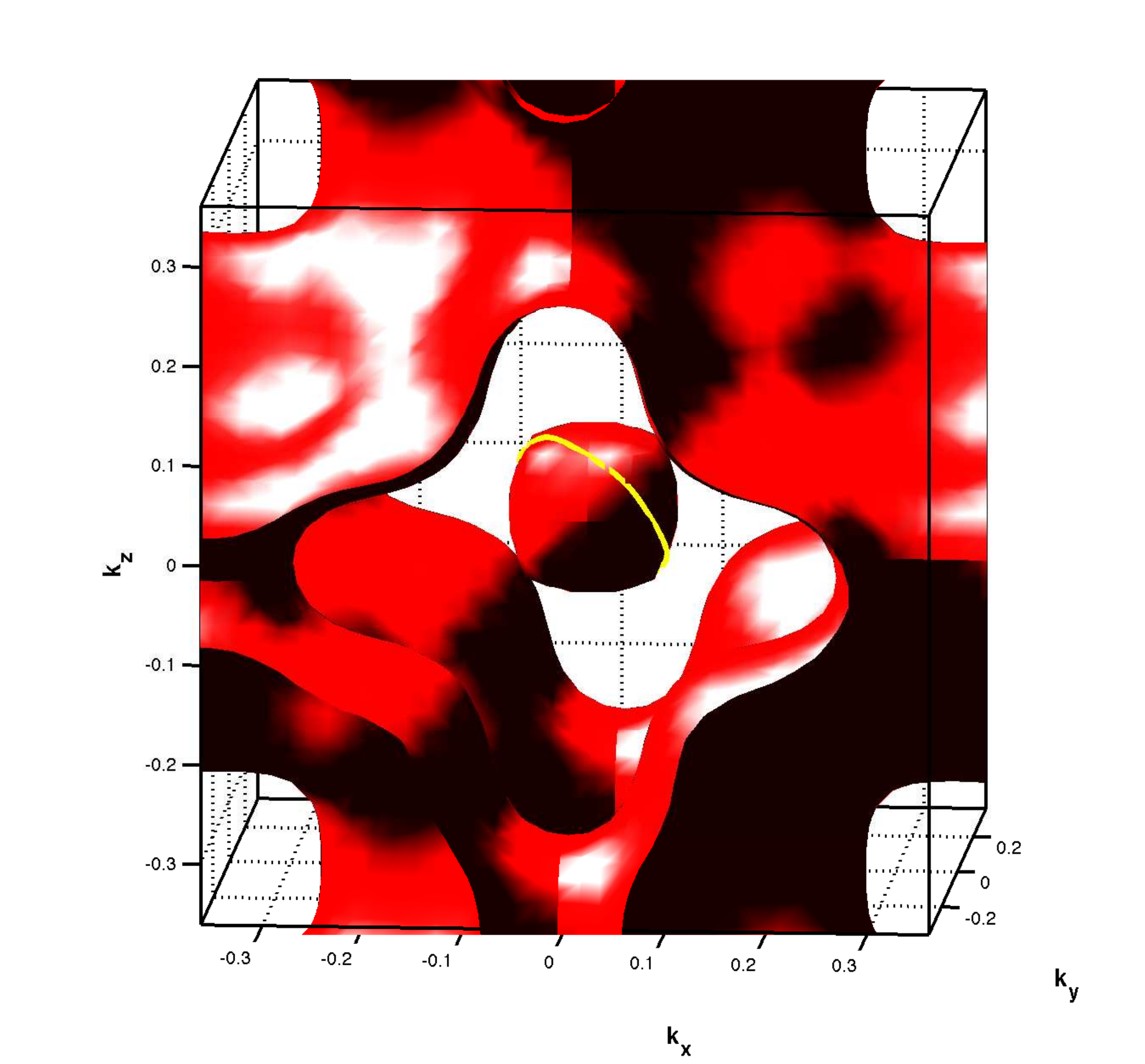}
\caption{(Color online)  Fermi surface from Band-1 in PM PuIn$_{\rm 3}$ compound.}
\label{fig:PuIn3PM_FS1}
\end{figure}

\begin{figure}
\includegraphics[scale=0.45]{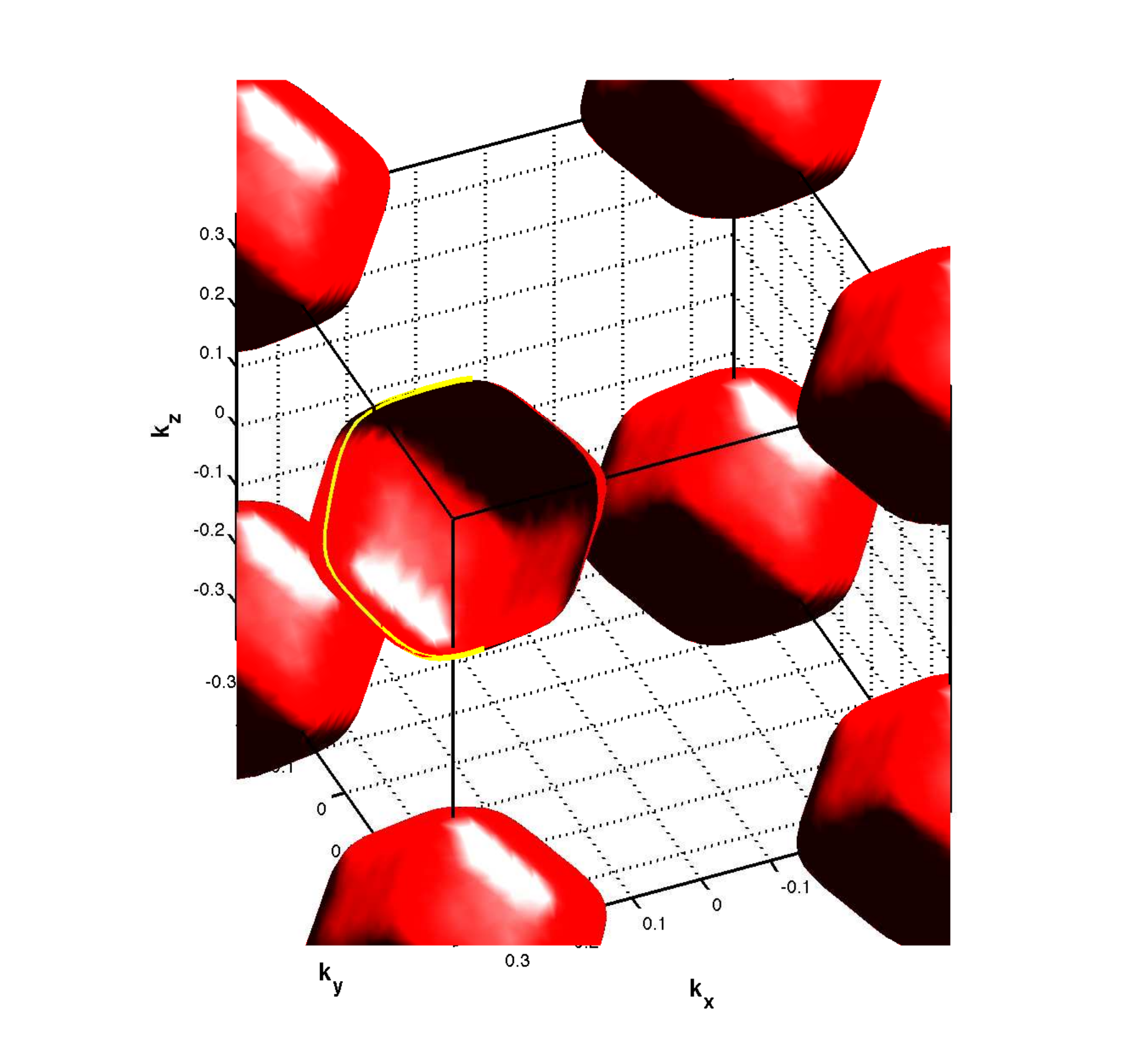}
\caption{(Color online) Fermi surface from Band-2 in PM PuIn$_{\rm 3}$ compound.}
\label{fig:PuIn3PM_FS2}
\end{figure}

\begin{figure}
\includegraphics[scale=0.55]{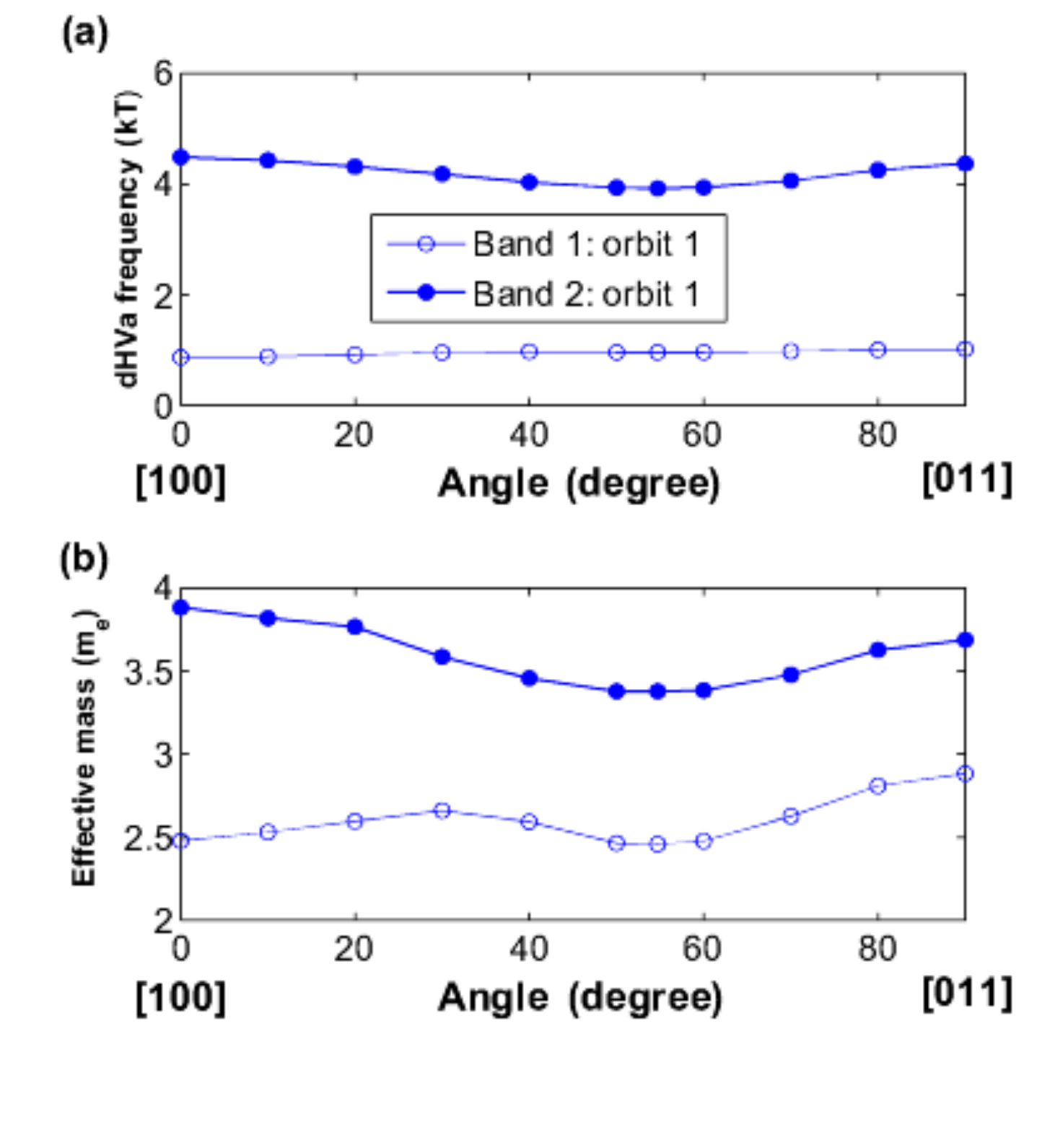}
\caption{(Color online) Angular dependence of dHvA frequencies and the average effective mass of the yellow orbits in PM PuIn$_{\rm 3}$ compound.
$\theta$ is the angle
spanned from the $[100]$  to $[011]$ orientation at the same plane. }
\label{fig:dHva2}
\end{figure}

The Fermi surfaces for PM PuIn$_{\rm 3}$ from two conduction bands are shown in Figs.~\ref{fig:PuIn3PM_FS1} and \ref{fig:PuIn3PM_FS2}.
The ordering of bands are based on the energies in increasing order in the band dispersion (Fig.~\ref{fig:FIG_PuIn3_PMBand}).
There are two bands across the Fermi level.
The characters of the dHvA orbits are tabulated  in Tables I and II respectively.
We present the dHvA orbits of interest by \dquote{yellow}  orbits or the label \dquote{orbit 1} for later discussion.
This is highlighted in the final two columns in Tables I and II.
For Band-1, when the magnetic field is chosen along [111] direction,
four types of dHvA orbits are identified in the BZ as listed in Table I. Three of the orbits are electron-like and one  is hole-like.
The dHvA orbit labeled by 1 originates from the Fermi surface near the center of the BZ  and has the dHvA frequency $F=0.9627~ {\rm kilo Tesla (kT)}$.
The location of the orbit 1 is shown in Fig.~\ref{fig:PuIn3PM_FS1} explicitly.
The other orbits with much larger dHvA frequencies are not our main focus since the dHvA frequencies are far from what have been observed
experimentally.~\cite{HAGA}
For Band-2,  only an electron-like (yellow) orbit with the dHvA frequency $F= 3.9206~{\rm kT}$  near the corner of the BZ is found, which is
shown in Fig.~\ref{fig:PuIn3PM_FS2}.
As far as the field angular dependence is concerned, we focus on the orbits which are connected with the same
Fermi surface surrounded  by the yellow orbits from Band-1
and Band-2 at the field orientation [111]. We observe the continuation of the orbits from the same Fermi surface spanned the whole angle range from [100] to [111] field orientation.
In addition, the calculated dHvA frequencies are not close to $2~{\rm kT}$ as shown in Fig.~\ref{fig:dHva2}(a).
Those facts contradict  the observed dHvA frequency (around $2~{\rm kT}$) which only survives a limited span of field angles.~\cite{HAGA}
This indicates that the assumption of the PM state in PuIn$_{\rm 3}$  is problematic for the interpretation of the electronic
properties of the PuIn$_{\rm 3}$ compound.
From the Fermi surface topology for PM PuIn3 in Figs.~9 and~10, we notice the nesting between the small Fermi surface at the $\Gamma$ point from Band-1 and the four Fermi surfaces at the corners of the BZ from Band-2. The nesting wave vector  between the Fermi surfaces is approximately $(\pi/a,\pi/a,\pi/a)$.
Therefore, magnetic instability for commensurate AFM order should indeed be favored.
In the following, we study commensurate AFM order in PuIn$_{\rm 3}$.

\begin{table}[htdp]
\caption{ dHvA orbits from Band-1: PM PuIn$_{\rm 3}$, ${\bf B} \parallel [111]$. h: holelike; e: electronlike.}
\begin{center}
\begin{tabular}{cccccc}
\hline
$F ({\rm kT})$ & $m^{*}(m_{e})$ & Type & Number & Color & Label \\
\hline
16.149  & 6.4320 & h & 1& -- & --\\
13.209 & 4.9218 & e & 1& -- & -- \\
 10.818 & 6.02 & e &2 & -- & --\\
 0.9627 & 2.4587 & e & 1 &  -- & 1\\
\hline
\end{tabular}
\end{center}
\label{table:1}
\caption{ The dHvA orbit from Band-2: PM PuIn$_{\rm 3}$, ${\bf B} \parallel [111]$. h: holelike; e: electronlike.}
\begin{center}
\begin{tabular}{cccccc}
\hline
$F({\rm kT})$  & $m^{*}(m_{e})$ & Type & Number of orbits & Color & Label \\
\hline
 3.9206 & 3.3774 & e & 1 & yellow & 1  \\
\hline
\end{tabular}
\end{center}
\label{table:2}
\end{table}

\subsubsection*{AFM PuIn$_{\rm 3}$ compound }

\begin{figure}
\includegraphics[scale=0.35]{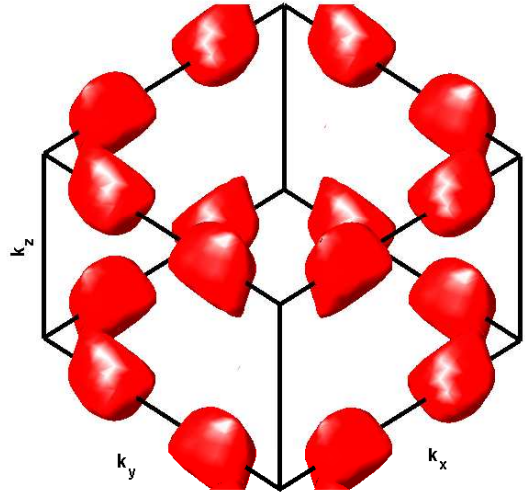}
\caption{(Color online)  Fermi surface from Band-1 in  AFM PuIn$_{\rm 3}$ compounds with the local spin polarization $\vec \sigma \parallel [001]$.}
\label{fig:FIG_PuIn3_AFMFS1}
\end{figure}

\begin{figure}
\includegraphics[scale=0.35]{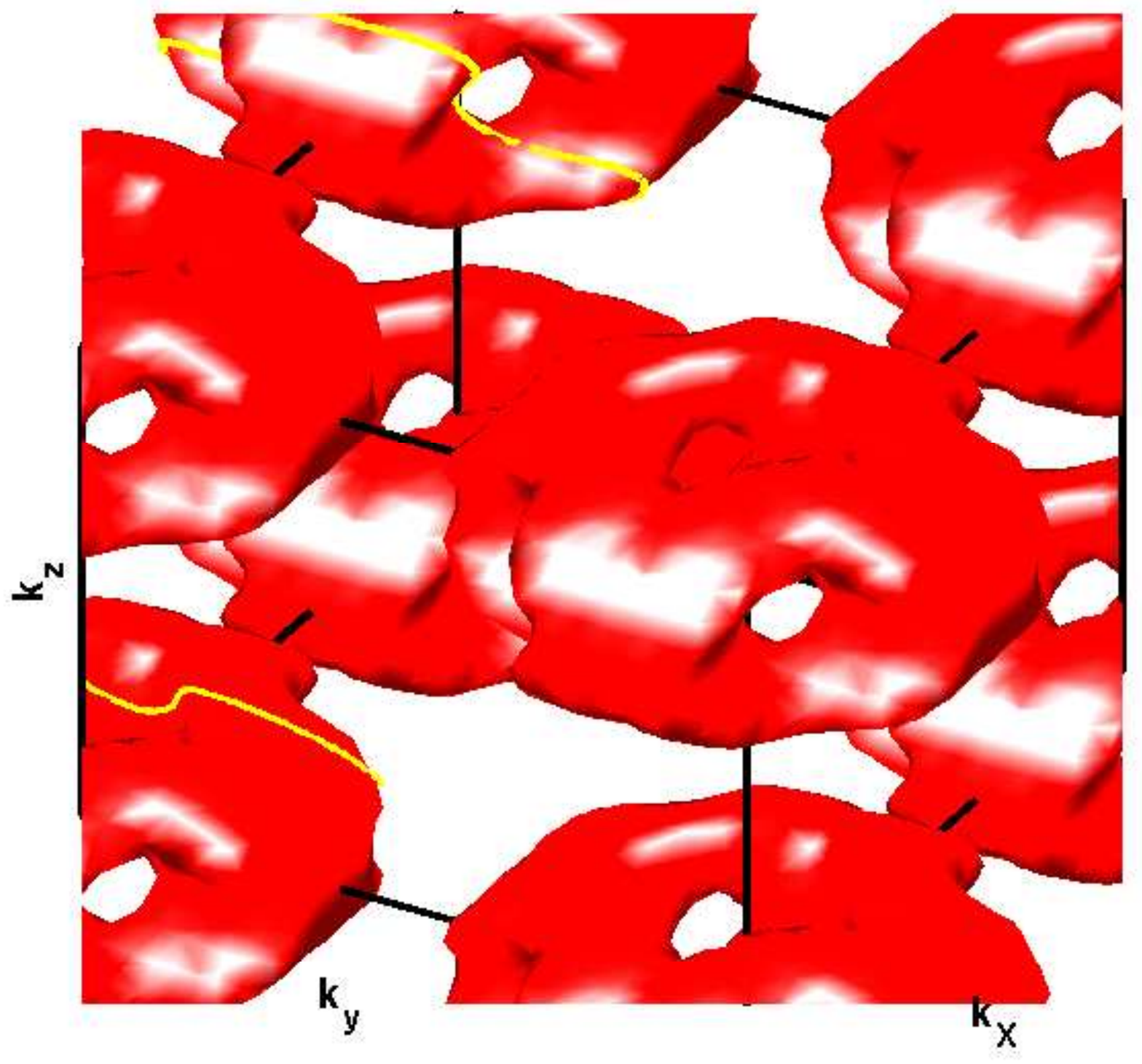}
\caption{(Color online) Fermi surface from Band-2 in AFM PuIn$_{\rm 3}$ compounds with the local spin polarization $\vec \sigma \parallel [001]$.}
\label{fig:FIG_PuIn3_AFMFS2}
\end{figure}

\begin{figure}
\includegraphics[scale=0.35]{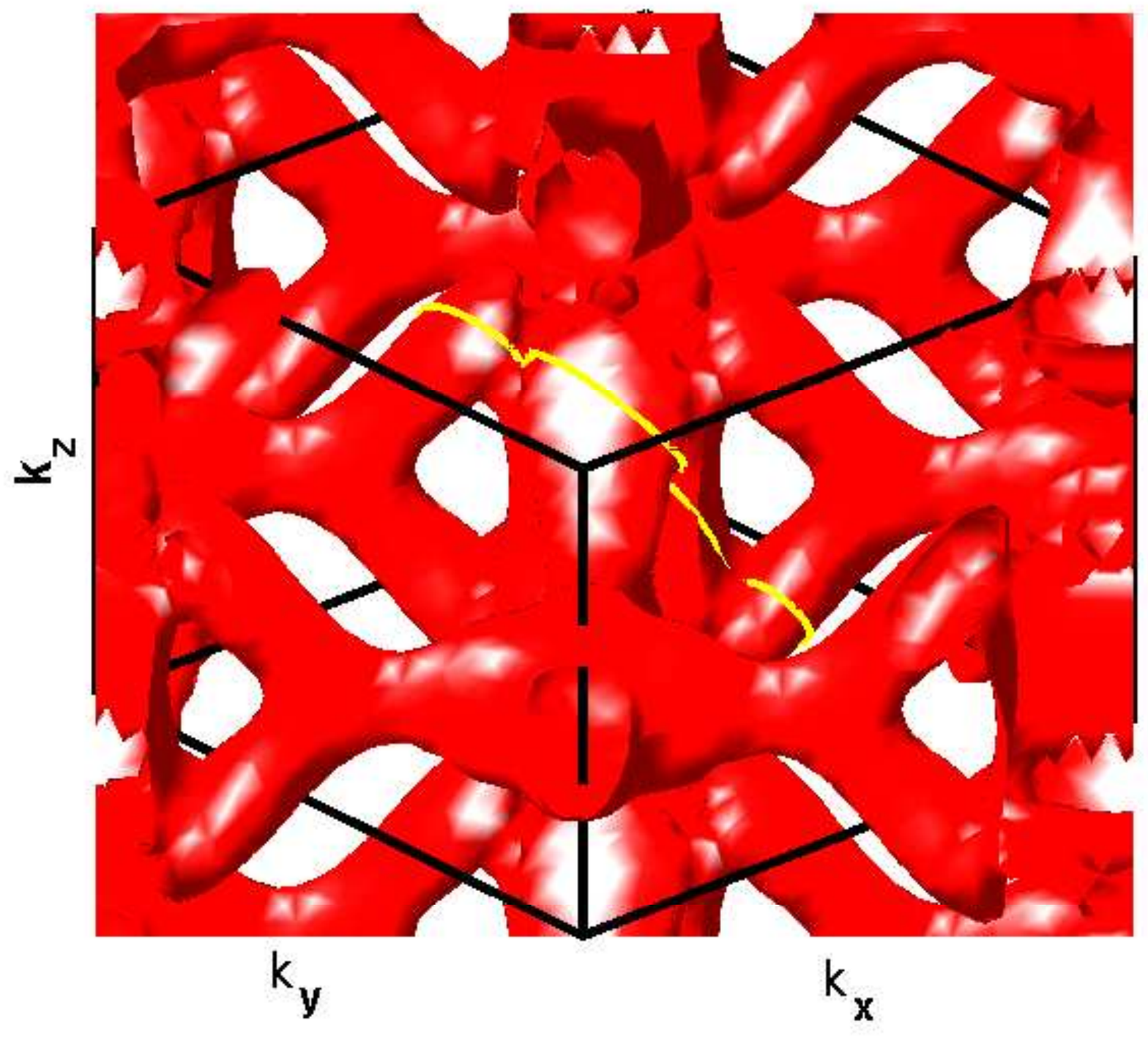}
\caption{(Color online) Fermi surface from Band-3 in AFM PuIn$_{\rm 3}$ compounds with the local spin polarization $\vec \sigma \parallel [001]$.}
\label{fig:FIG_PuIn3_AFMFS3}
\end{figure}

\begin{figure}
\includegraphics[scale=0.35]{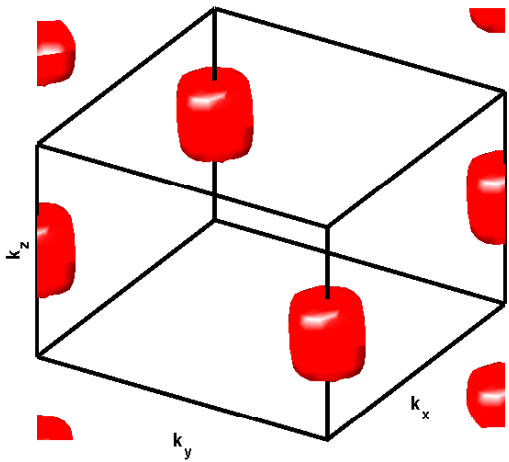}
\caption{(Color online) Fermi surface from Band-4 in AFM PuIn$_{\rm 3}$ compounds with the local spin polarization $\vec \sigma \parallel [001]$.}
\label{fig:FIG_PuIn3_AFMFS4}
\end{figure}

From the band structure calculations as shown earlier in Fig.~\ref{fig:FIG_PuIn3_AFMBand},
there are $4$ bands across the Fermi level for AFM PuIn$_{3}$.
The labeling of bands still follow the band energies (Band-1, Band-2, etc.).
The corresponding Fermi surfaces and dHvA orbits (when the external magnetic field is orientated along $[1 1 1]$ direction) are
illustrated in Figs.~\ref{fig:FIG_PuIn3_AFMFS1}-\ref{fig:FIG_PuIn3_AFMFS4}, respectively.
We are interested in dHvA orbits with dHvA frequencies around $2~{\rm kT}$, which are labeled as orbit 1 for Band-2 and Band-3 in
Tables  IV and V. The information of other dHvA orbits due to Band-1 and Band-4 are listed in Tables III and VI.
We provide the Fermi surface topology for Band-1 and Band-4 in Figs.~\ref{fig:FIG_PuIn3_AFMFS1} and \ref{fig:FIG_PuIn3_AFMFS4} for
reference.

\begin{figure}
\includegraphics[scale=0.55]{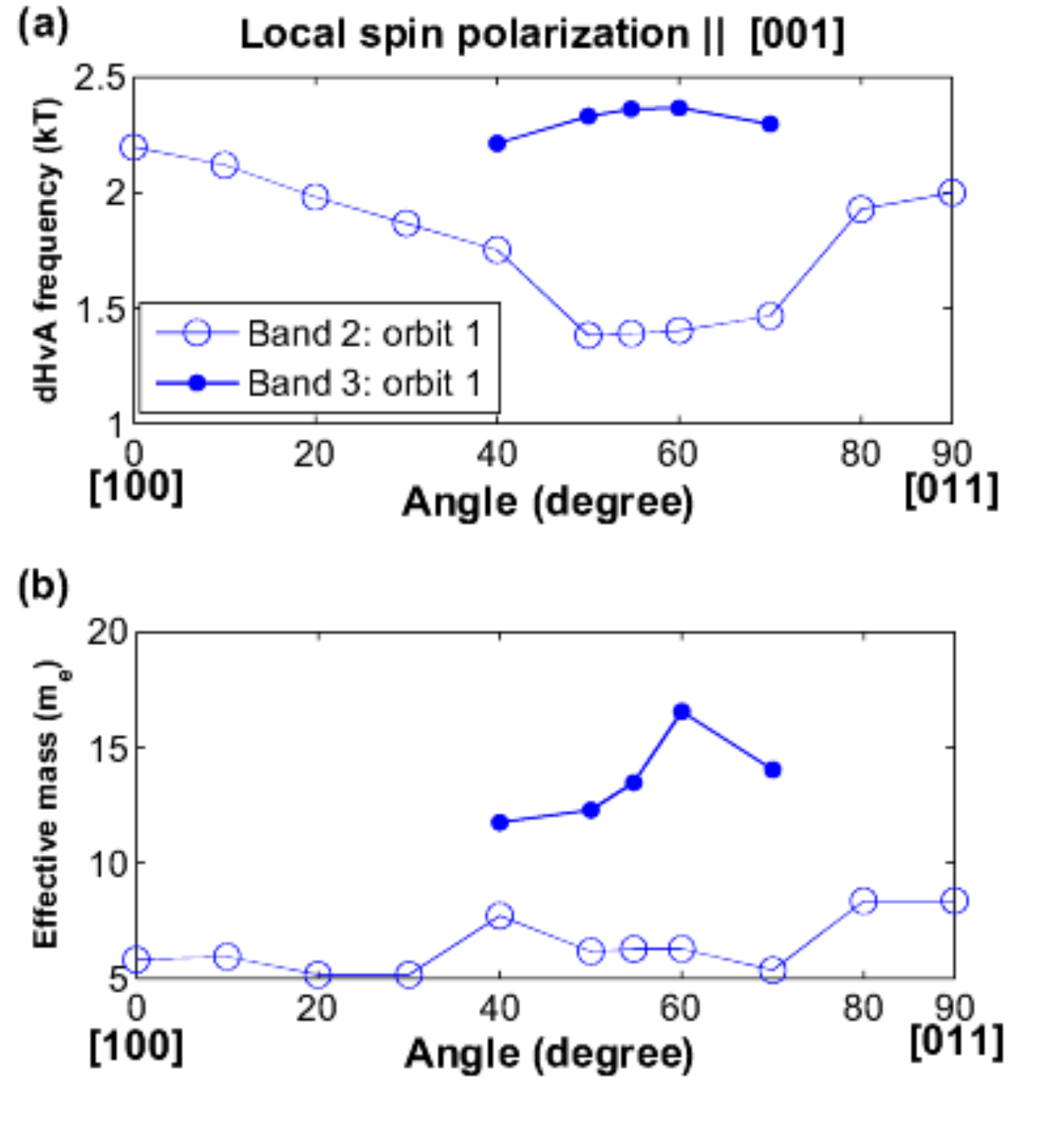}
\caption{(Color online) Angular dependence of dHvA frequencies and effective mass of chosen yellow orbits for AFM PuIn$_{\rm 3}$ compound. }
\label{fig:dHva3}
\end{figure}

For Band-2, the yellow orbits with the dHvA frequency of  $1.3908~{\rm kT}$  around the Fermi surface
at the corner of the BZ are illustrated in Fig.~\ref{fig:FIG_PuIn3_AFMFS2}.
When the field orientation varies, one expect dHvA orbits at different angle are continued orbits from the same Fermi surface.
Therefore, we expect the dHvA orbits inherited from the same Fermi surface  span the whole range of the angle as shown in Fig.~\ref{fig:dHva3}.
The dHvA frequencies are lowest around [111] orientation (angle$=54.7356^{o}$) and the value is around $1.5~{\rm kT}$.
The effective band mass is around $5m_{e}$ where $m_{e}$ is the free electron mass.
For band-3, in Fig.~\ref{fig:FIG_PuIn3_AFMFS3}, the dHvA frequencies around $2.3~{\rm kT}$ spanned through a limited range of angles
around [111]  as can be understood by the shape of the corresponding Fermi surface.  The dHvA frequencies have a local maxima around $[111]$ orientation instead.
The effective mass is above $10 m_{e}$ .
Even though we have a good agreement with the experiment~\cite{HAGA} on the dHvA frequencies with a limited angle span,
the predicted dHvA frequencies have a local maxima instead of a minimum as observed experimentally.
There is uncertainty in assuming magnetization orientation of AFM order along [001] orientation.
We expect the orientation of the magnetic order will lead to qualitative effects on the Fermi surface topology because of
the strong spin-orbital coupling effects in PuIn$_{\rm 3}$ compounds.

\begin{table}
\caption{ The dHvA orbit from Band-1: AFM PuIn$_{\rm 3}$, ${\bf B} \parallel [111]$, ${\vec \sigma}  \parallel [001]$. h: holelike; e: electronlike.}
\begin{center}
\begin{tabular}{cccc}
\hline
$F$ (${\rm kT}$) & $m^{*}(m_{e})$ & Type & Number  of orbits \\
\hline
0.4424 & 2.1654 & h & 2 \\
0.4412  & 2.3620 & h & 2 \\
\hline
\end{tabular}
\end{center}
\label{default}
\caption{ dHvA orbits from Band-2: AFM PuIn$_{\rm 3}$, ${\bf B} \parallel [111]$, ${\vec \sigma} \parallel [001]$. h: holelike; e: electronlike.}
\begin{center}
\begin{tabular}{cccccc}
\hline
$F$ (${\rm kT}$) & $m^{*}(m_{e})$ & Type & Number of orbits & Color & Label  \\
\hline
  1.3908&6.2948  & h & 2& yellow & 1  \\
   1.006& 3.7899  & h & 2& -- & --\\
   0.5189& 2.6069  & h &  2 &   -- & -- \\
   0.3335& 2.2621  & h &  4 &   -- & --\\
\hline
\end{tabular}
\end{center}
\label{default}
\caption{ dHvA orbits from  Band-3: AFM PuIn$_{\rm 3}$, ${\bf B} \parallel [111]$, ${\vec \sigma} \parallel [001]$. h: holelike; e: electronlike.}
\begin{center}
\begin{tabular}{cccccc}
\hline
$F$ (${\rm kT}$) & $m^{*}(m_{e})$ & Type & Number of orbits &Color & Label \\
\hline
   2.3615& 13.4852  & e & 1 & yellow & 1 \\
  0.7796& 1.8097  & e & 2  & -- & --\\
  0.6833 & 3.0840  & e & 2 & -- &--  \\
   0.577& 11.017 & e & 2 & --&--\\
   0.2279&2.2562 & e & 4 & --&--\\
   0.0386&0.8525&e&2&--&--\\
\hline
\end{tabular}
\end{center}
\label{default}
\caption{ The dHvA orbit from Band-4: AFM PuIn$_{\rm 3}$, ${\bf B} \parallel [111]$, ${\vec \sigma} \parallel [001]$. h: holelike; e: electronlike.}
\begin{center}
\begin{tabular}{cccc}
\hline
$F$ (${\rm kT}$) & $m^{*}(m_{e})$ & Type & Number of orbits  \\
\hline
0.5277  & 1.0592 & e & 1 \\
\hline
\end{tabular}
\end{center}
\label{default}
\end{table}

\begin{figure}
\includegraphics[scale=0.35]{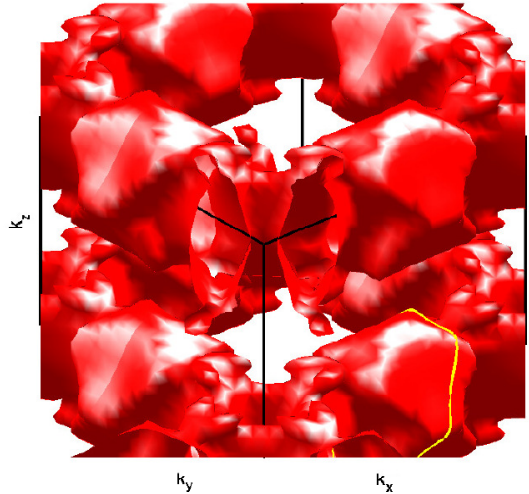}
\caption{(Color online)  Fermi surface from Band-1 in  AFM PuIn$_{\rm 3}$ compounds with the spin polarization $\vec \sigma \parallel [101]$. h: holelike; e: electronlike.}
\label{fig:PuIn3AFM111_FS1}
\end{figure}

\begin{figure}
\includegraphics[scale=0.35]{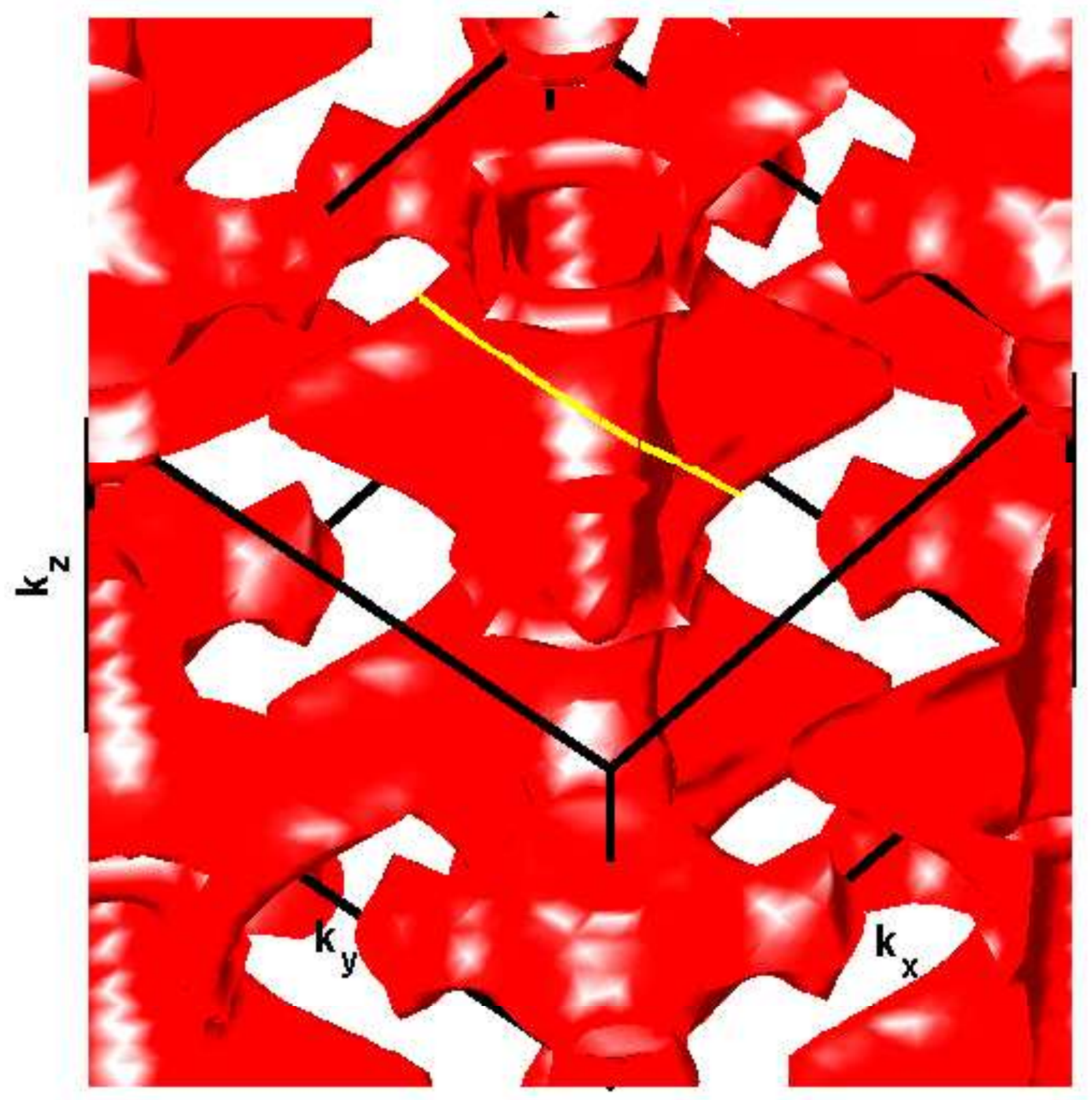}
\caption{(Color online) Fermi surface from Band-2 in AFM PuIn$_{\rm 3}$ compounds with the spin polarization $\vec \sigma \parallel [101]$. h: holelike; e: electronlike.}
\label{fig:PuIn3AFM111_FS2}
\end{figure}

\begin{figure}
\includegraphics[scale=0.35]{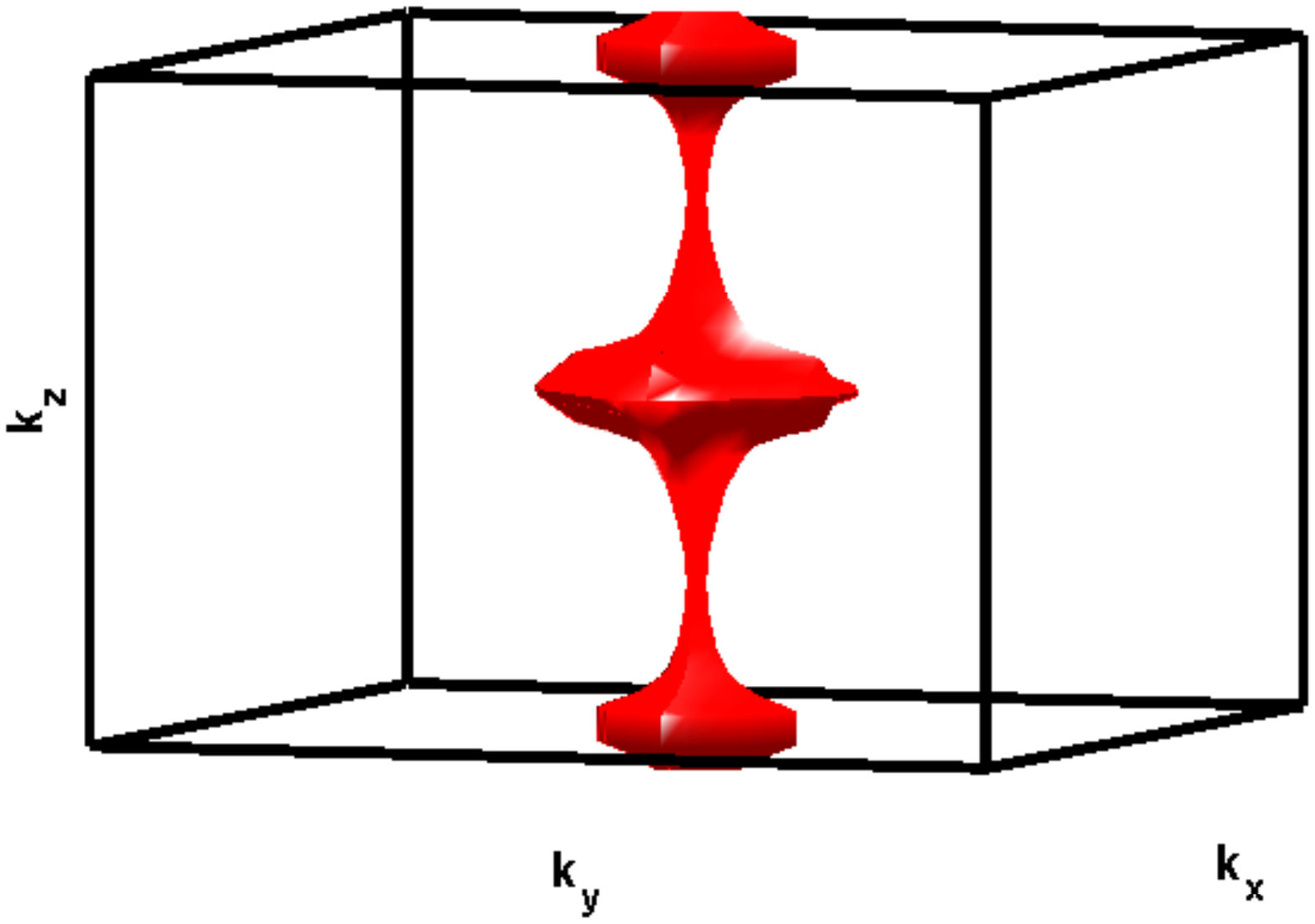}
\caption{(Color online) Fermi surface from Band-3 in AFM PuIn$_{\rm 3}$ compounds with the spin polarization $\vec \sigma \parallel [101]$. h: holelike; e: electronlike.}
\label{fig:PuIn3AFM111_FS3}
\end{figure}

\begin{figure}
\includegraphics[scale=0.55]{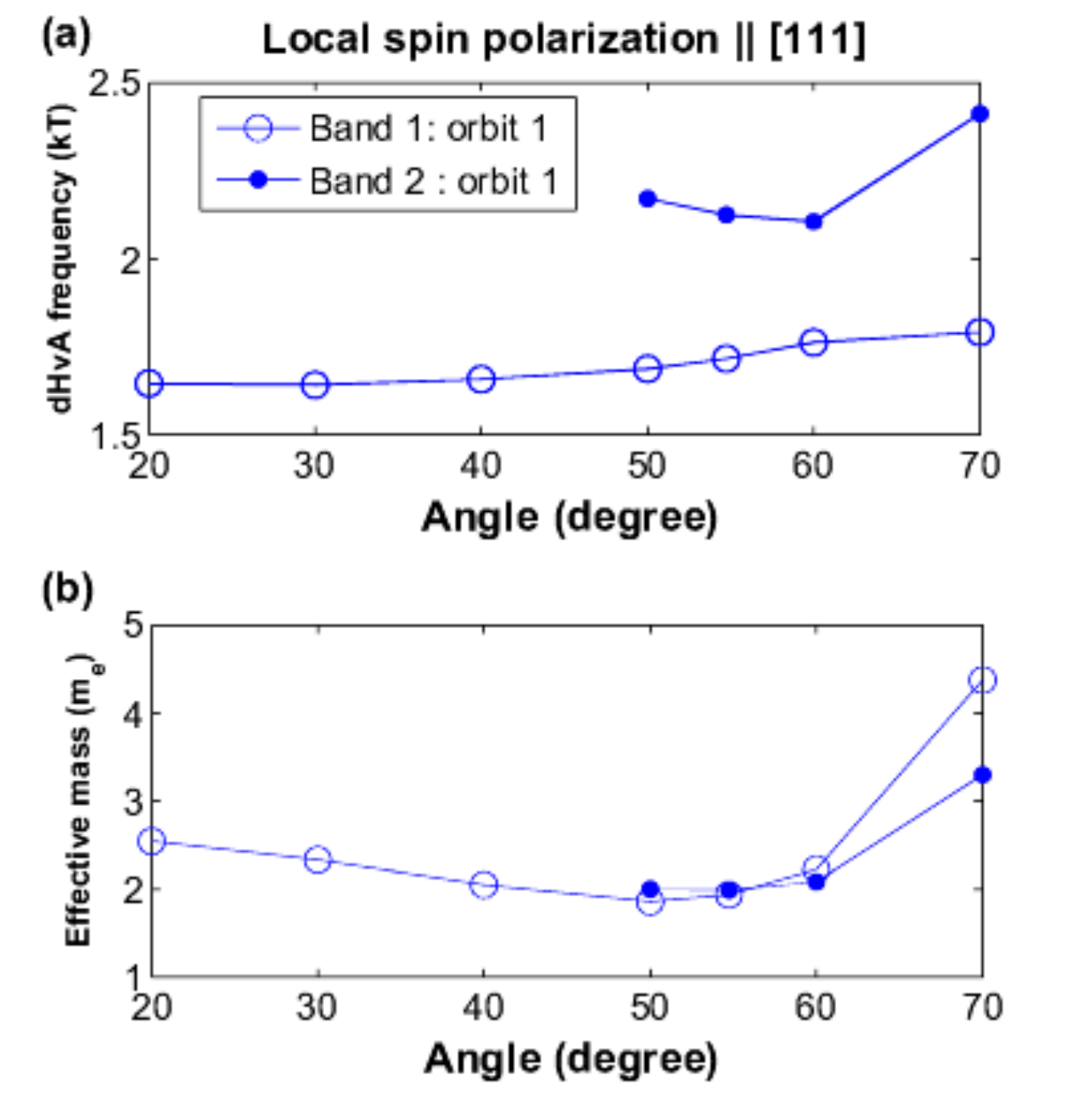}
\caption{(Color online) Angular dependence of dHvA frequencies and effective mass of the chosen yellow orbits for AFM PuIn$_{\rm 3}$ compound.
Notice that the predicted dHvA frequencies for Band 2 only exist between 50 and 70 degrees as shown by the filled circles. }
\label{fig:dHva4}
\end{figure}

Therefore, we have also considered an easy axis of magnetization along [101] direction.
In this case, we predict three $5f$ bands across the Fermi level.
We show the Fermi surface topology corresponding to three bands across the Fermi level in
Figs.~\ref{fig:PuIn3AFM111_FS1}--\ref{fig:PuIn3AFM111_FS3}.
The details of the dHvA orbits along [111] orientation are shown in Tables VII, VIII, and IX.
We identify an orbit with dHvA frequency  $F=2.125~{\rm kT}$  marked by a yellow trace in the Fermi surface around $\Gamma$ point from Band-2 in
Fig.~\ref{fig:PuIn3AFM111_FS2}. By the calculated angular dependence of the dHvA orbits for Band-2 in Fig.~\ref{fig:dHva4}(a), we identify the dHvA orbits
are limited to certain angle range around [111]  orientation. In addition, the effective mass has a local minima around [111] orientation
as suggested by experiment.~\cite{HAGA}  However, the effective mass is only about $2m_{e}$. For the dual nature of
the $5f$ electrons, the temporal quantum fluctuations can renormalize the effective band mass dramatically. We do not expect
density-functional theory as presented in this paper can capture this effect.

As far as the calculated DOS is concerned in comparison to experimental photoemission measurements,~\cite{Joyce}
we find a good agreement at the Fermi level with a noticeable DOS (see Fig.~21) as opposed to the small DOS in PM PuIn3 (see Fig.~6).
In addition, we also see
a main feature below the Fermi surface around -5eV due to delocalized bands. However, by our GGA calculation,
we observe a much sharper spectral feature
between $E-E_{F}=-2eV$ and the Fermi level. This is due to the failure of GGA to treat localized strong correlation
in Pu f orbitals. The strong correlation is expected to broaden the spectral features and renormalize the DOS obtained by GGA closer to
experimental observation. This effect should be captured by other more advanced approaches to treat  the strong correlation, such as dynamical mean-field theory(DMFT).~\cite{DMFT}

\begin{table}
\caption{ The dHvA orbit from Band-1: AFM PuIn$_{\rm 3}$, ${\bf B} \parallel {\vec \sigma} \parallel [101]$. h: holelike; e: electronlike.}
\begin{center}
\begin{tabular}{cccccc}
\hline
$F$ (${\rm kT}$) & $m^{*} (m_{e})$ & Type & Number  of orbits & Color & Label  \\
\hline
3.2569&  5.8858 & h & 1& -- & --  \\
 1.7165  & 1.9391  & h &1 & yellow & 1\\
   0.9876 & 3.4407   & h & 2  &   -- & -- \\
   0.9697 & 2.8944  & h &  2 &   -- & --\\
   0.8421 & 3.5473& h & 2& --& -- \\
   \hline
\end{tabular}
\end{center}
\label{default}
\caption{ dHvA orbits from Band-2: AFM PuIn$_{\rm 3}$, ${\bf B} \parallel {\vec \sigma} \parallel [101]$. h: holelike; e: electronlike.}
\begin{center}
\begin{tabular}{cccccc}
\hline
$F$ (${\rm kT}$) & $m^{*} (m_{e})$ & Type & Number of orbits & Color & Label  \\
\hline
  3.7022&6.2443 &  e& 1& -- & --  \\
  2.1250 & 1.9877  &e  & 1& yellow& 1\\
   1.2687& 2.8742  & e & 2  &   -- & -- \\
  1.1595 & 1.5520   & e &   1&   -- & --\\
  1.0834 & 1.7674  & e &   2&   -- & --\\
   \hline
\end{tabular}
\end{center}
\label{default}
\caption{ dHvA orbits from  Band-3: AFM PuIn$_{\rm 3}$, ${\bf B}\parallel {\vec \sigma}\parallel [101]$. h: holelike; e: electronlike.}
\begin{center}
\begin{tabular}{cccc}
\hline
$F$ (${\rm kT}$) & $m^{*} (m_{e})$ & Type & Number of orbits \\
\hline
   0.3359 &2.3248   &  e& 1  \\
   0.3358 &2.4819   &  e& 1  \\
   0.3056& 2.1990  & e &  1 \\
  0.1935&  1.7949 & e &  1  \\
\hline
\end{tabular}
\end{center}
\label{default}
\end{table}

\begin{figure}
\includegraphics[scale=0.38]{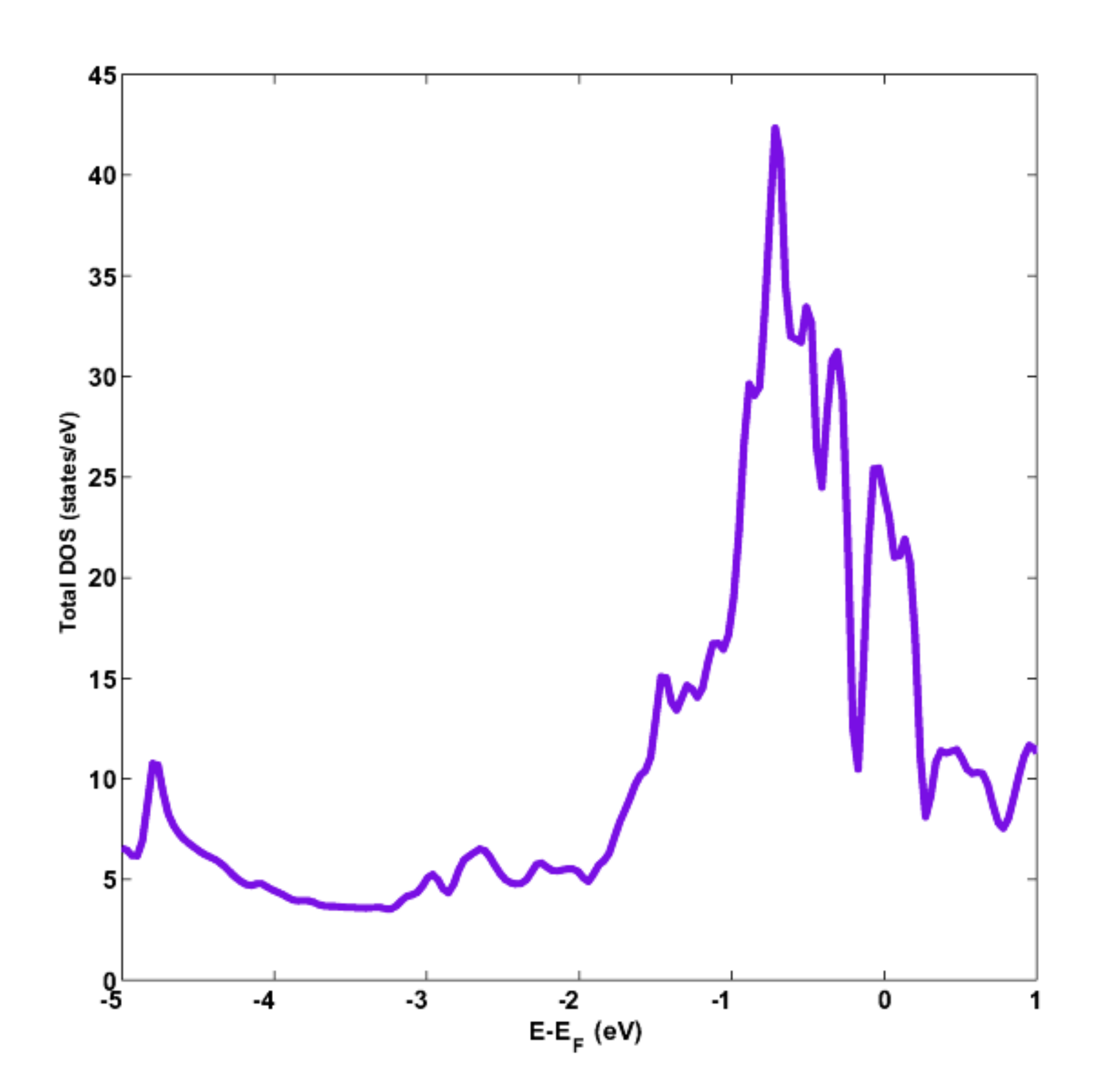}
\caption{(Color online) Total DOS in AFM PuIn$_{\rm 3}$ compounds with the spin polarization $\vec \sigma \parallel [101]$.}
\label{fig:111DOS}
\end{figure}

\subsubsection*{PM PuSn$_{\rm 3}$ compound}
\begin{figure}
\includegraphics[scale=0.32]{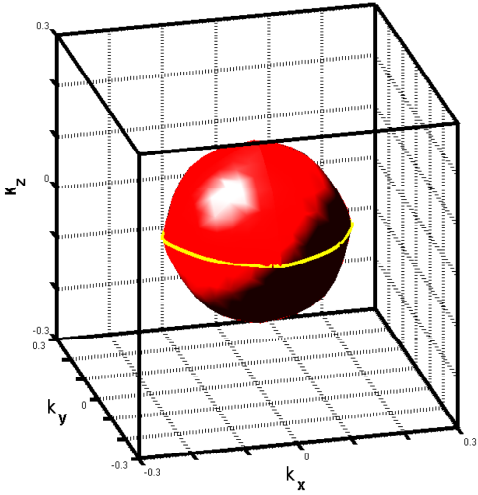}
\caption{(Color online) Fermi surface from Band-1 in PM PuSn$_{\rm 3}$ compound. }
\label{fig:FIG_PuSn3_PMFS1}
\end{figure}

\begin{figure}
\includegraphics[scale=0.32]{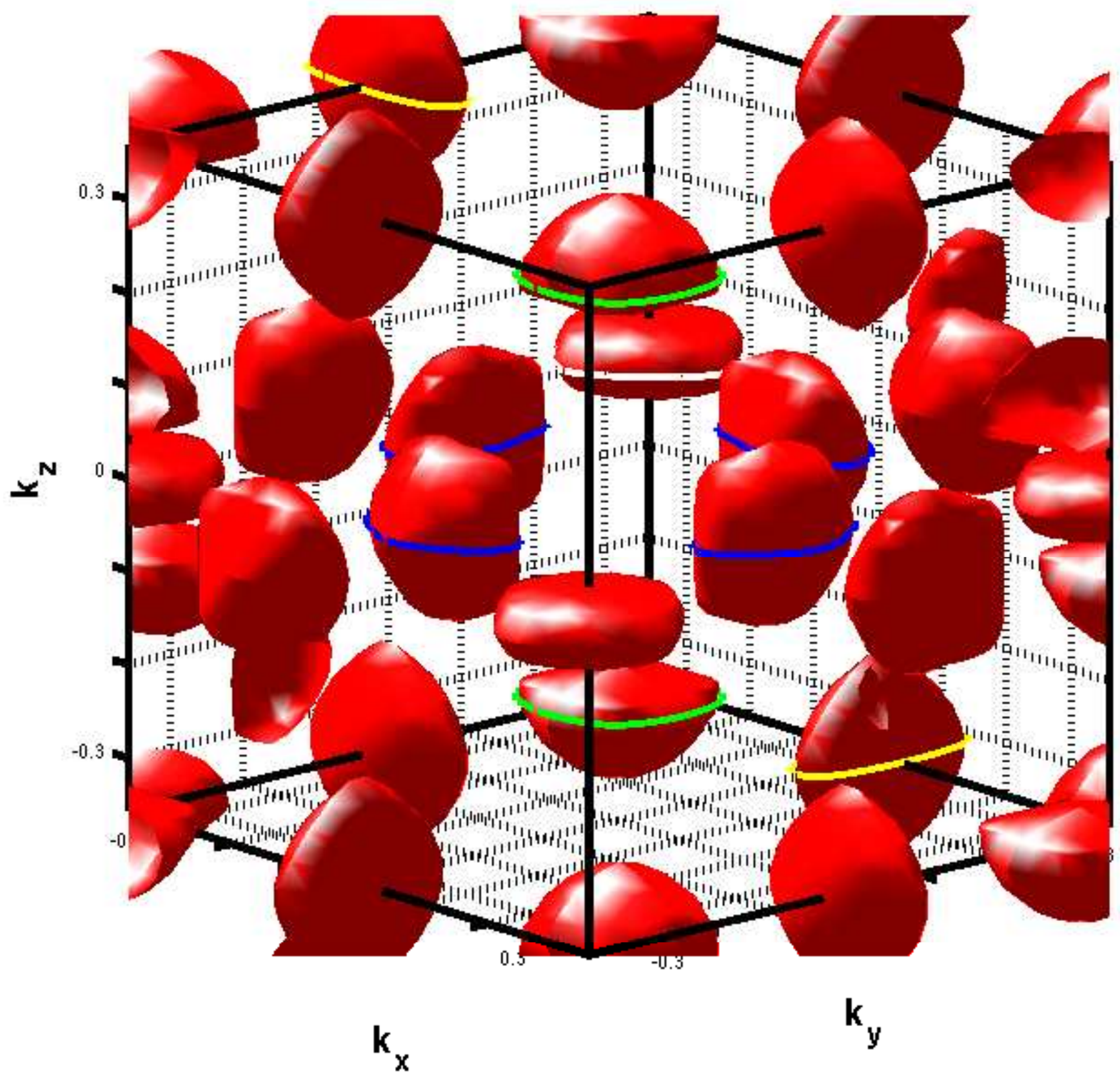}
\caption{(Color online) Fermi surface from Band-2 in PM PuSn$_{\rm 3}$ compound. }
\label{fig:FIG_PuSn3_PMFS2}
\end{figure}

So far, dHvA effects have not yet been successfully observed  in PuSn$_{\rm 3}$ compounds.
By our calculation, we find multiple dHvA orbits in a narrow range of frequencies at the field orientation [111].  This may cause complications
on observing clean signals. However, we do find a simple Fermi surface topology at the field orientation [001], which can be easier
to observe and thus is the case we focus on in this section.
With the PuSn$_{\rm 3}$ band structure in Fig.~\ref{fig:FIG_PuSn3_Band}, we observe two energy bands crossing the Fermi energy $E_{F}$, which only have similar
dispersion as PM PuIn$_{\rm 3}$ near the point $\Gamma=(0,0,0)$ in the BZ. The cutting points at the Fermi energy in Fig.~\ref{fig:FIG_PuSn3_Band}  are the projection of the three dimensional Fermi surface plot along the line between symmetry points in the BZ.
We show the Fermi surfaces for those two bands in Figs.~\ref{fig:FIG_PuSn3_PMFS1} and~\ref{fig:FIG_PuSn3_PMFS2}, respectively.
Our calculation shows  one hole-like (yellow) orbit (with $F=3.273~{\rm kT}$ ) near the $\Gamma$ point in the BZ, which agrees with the band dispersion at Band-1 near the $\Gamma$ point.
For Band-2, the four dHvA orbits are identified and all of them are hole-like at the field orientation $[001]$ (See Table VIII), denoted by colored closed curves in Fig.~\ref{fig:FIG_PuSn3_PMFS2}.
The maximum dHvA frequency is given by the white cyclotron orbit with $F=1.312~{\rm kT}$ near the M point in the BZ.
By observing the Fermi points and the band dispersion nearby for Band-2 in Fig.~\ref{fig:FIG_PuSn3_Band}, one can see the agreement with the Fermi surface.
In addition, from the Fermi surface topology, one can see there is no clear evidence for a dominant Fermi surface nesting. This might explain the absence of magnetism
in PuSn$_{\rm 3}$.~\cite{PuSn3PM}

\begin{figure}
\includegraphics[scale=0.55]{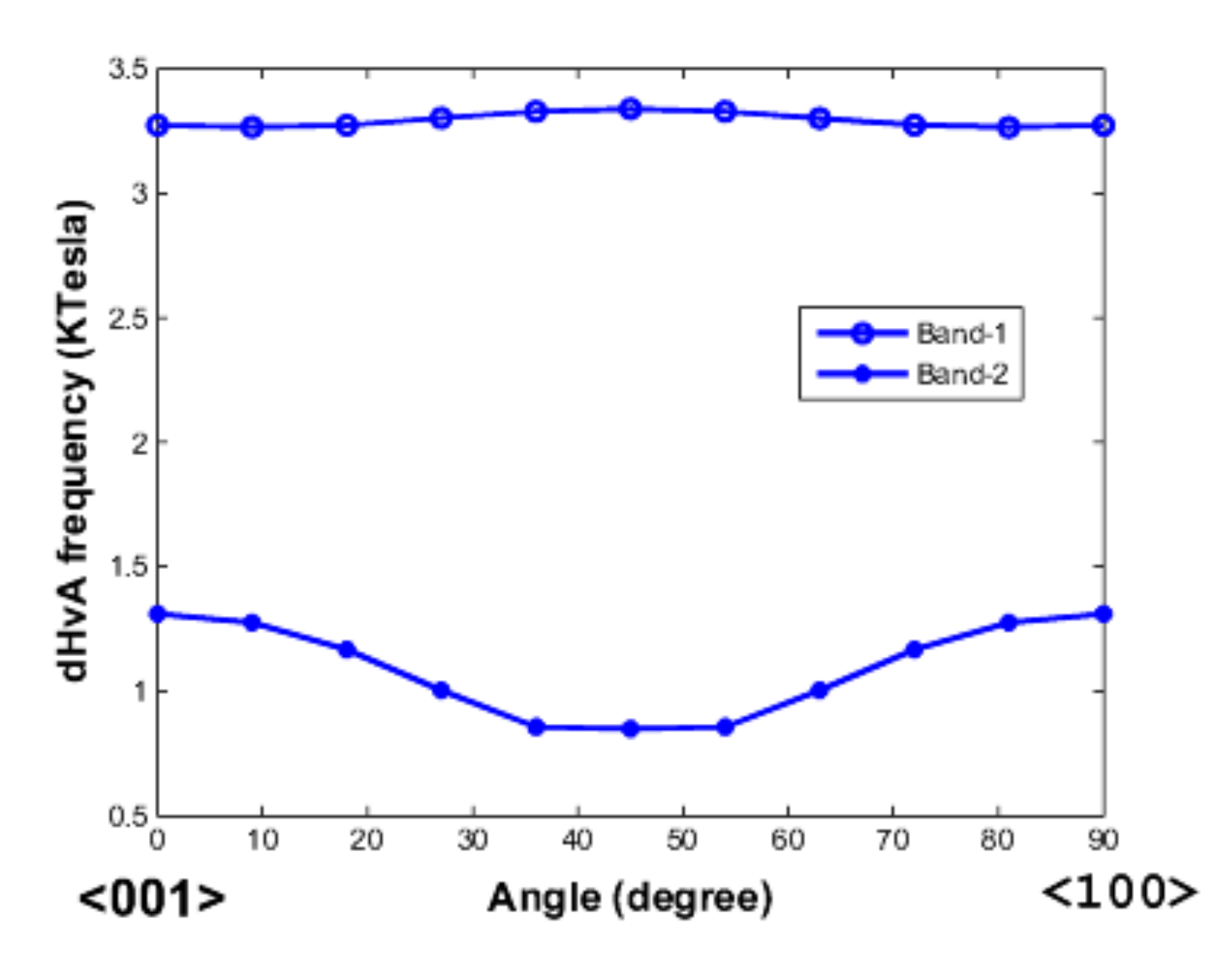}
\caption{(Color online) Angular dependence of dHvA frequencies in PM PuSn$_{\rm 3}$ compound.
The dHvA orbit of Band-1 is inherent from the same Fermi surfaces circled by the yellow orbit
in Fig.~17.
The dHvA orbit of Band-2 at different angle is inherent from the same Fermi surfaces circled by the green orbits
in Fig.~18.}
\label{fig:dHva1}
\end{figure}

\begin{table}[htdp]
\caption{ The dHvA orbit from Band-1:  PM PuSn$_{\rm 3}$, ${\bf B} \parallel [001]$. h: holelike; e: electronlike.}
\begin{center}
\begin{tabular}{ccccc}
\hline
$F$ (kT) & $m^{*} (m_{e})$ & Type & Number of orbits &Color \\
\hline
3.273  & 1.5976 & h & 1& yellow\\
\hline
\end{tabular}
\end{center}
\label{default}
\caption{ dHvA orbits from Band-2:  PM PuSn$_{\rm 3}$, ${\bf B} \parallel [001]$. h: holelike; e: electronlike.}
\begin{center}
\begin{tabular}{ccccc}
\hline
$F$ (kT) & $m^{*} (m_{e})$ & Type & Number of orbits &Color \\
\hline
 0.5550 & 0.4749 & h & 2& yellow  \\
 0.6873 & 0.7955 & h & 4& blue \\
 1.230   & 1.6842 & h &2 & green\\
 1.312 & 0.8721 & h &  1 &   white \\
\hline
\end{tabular}
\end{center}
\label{default}
\end{table}
In Fig.~\ref{fig:dHva1}, we show the field angle dependence of dHvA frequencies for the yellow dHvA orbit in Band-1 and the white dHvA orbit
in Band-2. The magnetic field varies from the crystal orientation $[100]$ to $[001]$ in PuSn$_{3}$ compound.
We observe that the dHvA frequencies are symmetrical with respect to the orientation at the angle $45^{o}$.
The dHvA frequency for Band-1 is insensitive to the orientation because of the corresponding spherical Fermi surface near the $\Gamma$ point and
is well separated in frequencies from the orbits due to Band-2.
However, the Fermi surface circled by the white orbit in Band-2 is not spherical (shown in Fig.~\ref{fig:FIG_PuSn3_PMFS2}); therefore, a strong field angle dependence due to the Fermi surface is obtained. The nature of dHvA orbits are given in Tables X and XI respectively for band 1 and band 2.
\\
\section{Conclusions and discussions}
We study the electronic properties for compounds PuSn$_{\rm 3}$ and PuIn$_{\rm 3}$ within density functional theory under the GGA approximation.
Our calculations for AFM PuIn$_3$, with magnetization along [001] direction, show that the Fermi surfaces with heavy effective band mass  produce the dHvA frequencies near $2~{\rm kT}$, in quantitative agreement with experiments but the field-angle dependence of dHvA  frequencies around $[111]$  orientation ($\approx 54.74$ degree) cannot be reconciled with experiment. However, when the magnetization aligns along [101]  direction, we can identify  new orbits showing good agreement with experimental
measurement in dHvA frequencies. We point out the predicted effective band mass values can be further renormalized due to temporal quantum fluctuations
which cannot be captured properly with the density-functional theory.
Experimental investigations of the effective band mass of the dHvA orbits and the magnetic order are crucial to those issues.

In addition, we predict the band structure and dHvA frequencies for PuSn$_{\rm 3}$, which is known to be paramagnetic, for guiding the search of
dHvA effects in PuSn$_{\rm 3}$ inter-metallic compounds.
Experimental observation of these frequencies in the compound will help the search for quantum oscillations in $\delta$-phase of Pu. As a known fact, the GGA underestimates the electronic correlations in complex materials. The evaluation of these effects on electron mass and the Fermi surface in the compounds studied here requires a more rigorous but  sophisticated quantum many-body approach and goes beyond the scope of the present study.

Finally, we comment on the factors, which are pertinent to  the observation of  predicted dHvA orbits.
When the dHvA frequency is too low, the temperature required to observe the well-defined quantum oscillations has to be much lower than the energy scale set by the cyclotron frequency. It poses a challenge to the experimental observation. The other constraint is set by the sample size. When the dHvA frequency is too low, the corresponding cyclotron radius can be much larger than the sample size, making its observation difficult. In addition, for high frequency orbits, the sample quality is extremely important. Any scattering due to impurities can damp out the dHvA oscillations within the cycle of quantum oscillation. We believe the capability and design of the experiments limit what can be observed in reality.

\begin{acknowledgments}
We  thank  Eric D. Bauer, Ross D. McDonald, Matthias J. Graf, and P. M. C. Rourke
for helpful discussions. This work was supported by U.S. DOE  at LANL under Contract No. DE-AC52-06NA25396 and LANL LDRD-DR Program (C.-C.W. \& J.-X.Z.)
\end{acknowledgments}

\end{document}